\documentclass[aps,pre,twocolumn,superscriptaddress,showpacs,floatfix]{revtex4-1}
\usepackage{graphicx}
\usepackage{amssymb,amsmath,amsbsy,amsfonts,bm}
\usepackage{epsfig}
\usepackage{subfigure}
\usepackage{multirow}
\usepackage{tikz}
\usepackage{hyperref}
\usepackage{gensymb}
\usepackage{threeparttable}
\usepackage{bbold}

\usepackage{xcolor}
\usepackage{color, colortbl}
\usepackage[percent]{overpic}

\definecolor{LightCyan}{rgb}{0.88,1,1}
\definecolor{LightMagenta}{rgb}{1.        ,  0.8       ,  0.89803922}

\definecolor{darklavender}{rgb}{0.45, 0.31, 0.59}
\definecolor{amethyst}{rgb}{0.6, 0.4, 0.8}

\newcommand{\ofr}{(\mathbf{r})}
\newcommand{\Sofr}{$S \ofr$}
\newcommand{\figref}[1]{Fig.~\ref{#1}}
\newcommand{\Betrag}[1]{\left | #1 \right |}

\newcommand{\nofr}{$\mathbf{\hat{n}} \ofr$}

\newcommand{\tetr}{$\mathbf{\hat{t}}$}

\newcommand{\Qnem}{Q}
\newcommand{\Qtet}{q}

\newcommand{\rene}{\color{black}}
\newcommand{\reneN}{\color{black}}

\begin{document}
\title{Topology of Orientational Defects in Confined Smectic Liquid Crystals}

\author{Paul A. Monderkamp}
\affiliation{Institut f\"ur Theoretische Physik II: Weiche Materie, Heinrich-Heine-Universit\"at D\"usseldorf, Universit\"atsstra{\ss}e 1, 
40225 D\"usseldorf, Germany}
\author{Ren\'e Wittmann}
\email{rene.wittmann@hhu.de}
\affiliation{Institut f\"ur Theoretische Physik II: Weiche Materie, Heinrich-Heine-Universit\"at D\"usseldorf, Universit\"atsstra{\ss}e 1, 
40225 D\"usseldorf, Germany}
\author{Louis B.\ G.\ Cortes}
\affiliation{\mbox{School of Applied and Engineering Physics, Cornell University,~Ithaca,~NY~14853,~USA}}
\author{Dirk G.\ A.\ L.\ Aarts}
\affiliation{Department of Chemistry, Physical and Theoretical Chemistry Laboratory, University of Oxford, South Parks Road, Oxford OX1 3QZ, United Kingdom}
\author{Frank Smallenburg}
\affiliation{Laboratoire de Physique des Solides, CNRS, Universit\'e Paris-Saclay,  91405 Orsay, France}
\author{Hartmut L\"owen}
\affiliation{Institut f\"ur Theoretische Physik II: Weiche Materie, Heinrich-Heine-Universit\"at D\"usseldorf, Universit\"atsstra{\ss}e 1, 
40225 D\"usseldorf, Germany}
\begin{abstract}
We propose a general formalism to characterize orientational frustration of smectic liquid crystals in confinement by interpreting the emerging networks of grain boundaries as objects with a topological charge.
  In a formal idealization, this charge is distributed in pointlike units of quarter-integer magnitude,
 which we identify with tetratic disclinations
located at the end points and nodes.
   This coexisting nematic and tetratic order is analyzed with the help of
 extensive Monte Carlo simulations for a broad range of two-dimensional confining geometries as well as colloidal experiments,
showing how the observed defect networks can be universally reconstructed from simple building blocks.
We further find that the curvature of the confining wall determines the anchoring behavior of grain boundaries, such
 that the number of nodes in the emerging networks and the location of their end points can be
 tuned by changing the number and smoothness of corners, respectively.

\end{abstract} 

\maketitle
Topological defects are ubiquitous in ordered states of matter \cite{chaikin_lubensky_1995,toptherdef,topcosmdomains,topdeftwocompAP,superconduc} and thus also emerge as a characteristic feature of liquid crystals \cite{PhLiqCrys,SoftMPhys,nem_defects,QuantSelAss,elemBB,radzihovsky2020_gaugetheorySM}, which exhibit various degrees of positional and orientational ordering \cite{PhLiqCrys,teerakapibal2018_LCglass,bolhuis1997tracing,Mederos2014,PD_Rene,dussi2018,peters2020}.
{\rene Frustrated orientational order in nematic liquid crystals typically manifests itself in the form of} singular points or lines.
Defects of this type may spontaneously form and annihilate in bulk due 
to fluctuations \cite{crawford1996liquid}, 
external influences such as electromagnetic fields \cite{reznikov2000photoalignment,saturn_rings,SaturnRingDefects,cladis1987dynamics,link1996simultaneous}, changes in temperature \cite{microwrinklegrooves,BrMotDef} or active motion \cite{actNem,gia_num_fluc,chardac2021}.
In these processes, the defect strength, quantified by a half-integer topological charge {\rene $\Qnem$}, is subject to a universal conservation law \cite{algebTop,senyuk2013topological}.  
The formation of topological defects can further be triggered in a controlled manner through confining the particles \cite{lavrentovich1986phase,lavrentovich2010topological,real_defch,dammone2012,VirNemConfGeom,dzubiella2000topological,trukhina2008,varga2014hard,Jackson2017,rectangularConfinement,kralj2014,majumdar2016multistable,garlea2016finite,exp_studies,
yaochen2020,basurto2020,kil2020} or inserting an obstacle \cite{poulindirect,ruhwandl1997monte,andrienko2001computer,stabAndRew,ilnytskyi2014topological,schlotthauer2017,senyuk2016,coll_transp}. 
In this case, the precise type, number, and location of defects depend on the particular geometry \cite{araki_tanaka2013,campbell2014topological,garlea2019colloidal} and particle properties \cite{garlea2016finite,vananders2014,SemFlexNem,chiappini2019,revignas2020}, 
 due to a delicate balance between elastic distortions and surface anchoring. 

\begin{figure}[t]
\begin{center}
\includegraphics[width=\linewidth]{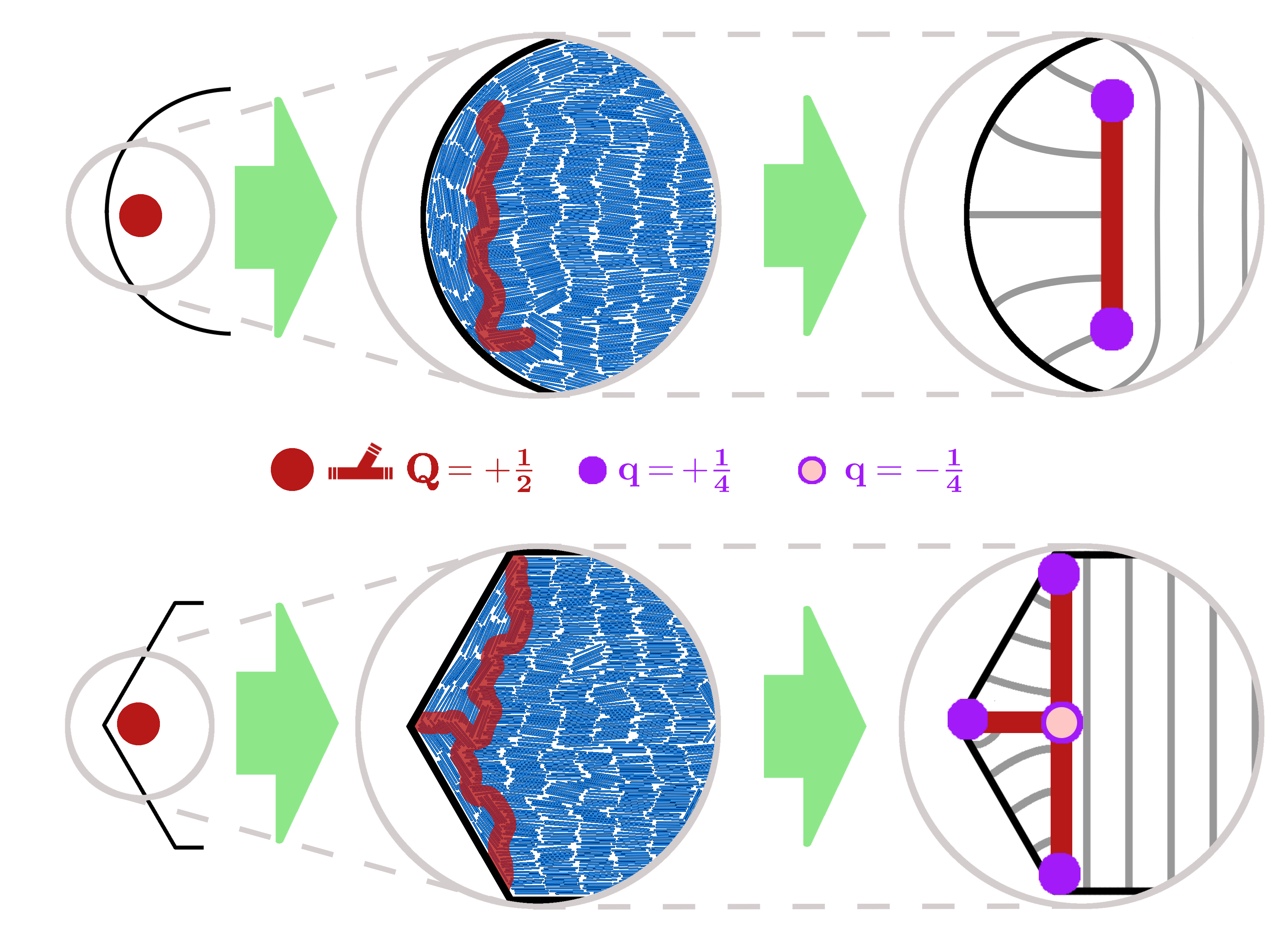}
\caption{
Topological characterization of  {\rene grain boundaries} in confined smectic liquid crystals. 
\textbf{Left:} coarse-grained topological structure with an idealized  {\rene nematic \reneN disclination} 
of half-integer charge {\rene $\Qnem$} induced by the presumed planar alignment with the nearby wall. 
\textbf{Middle:} particle-resolved simulation snapshots of hard rods with highlighted {\rene grain boundaries}
in the form of a line close to a circular wall (top row) and a network induced by corners (bottom row). 
\textbf{Right:} continuum model {\rene with quarter-integer tetratic point charges $\Qtet$}
at the end points and nodes, {\rene which can be interpreted as the distribution of the spatially extended charge $\Qnem$}. 
}
\label{fig_Lupenbild}
\end{center}
\end{figure}

The characteristic positional order of smectic liquid crystals 
breaks the symmetry of the homogeneous nematic phase and affects the elastic properties \cite{PhLiqCrys,lagerwallSM2006}.  
{\reneN The constraints associated with the layer structure \cite{poenaru1981,mosna2012}
stabilize distortions of the bulk smectic lattice \cite{chen2009symmetry,machon2019} 
which do not exist in nematic liquid crystals \cite{PhLiqCrys}.
These include purely positional defects called edge dislocations \cite{EdgeDisloc,SmEdgeDislogShear,NanStrucEdgeDisloc,kamien2016}
but also more complex objects like focal conic domains \cite{kleman2000grain,bramble2007,FocalDomains,liarte2015}.} 
{\rene In many cases, orientational frustration in smectic phases can be well described in terms of topological defects in the nematic order that is inherent to the symmetry of the smectic phase. However,} in the paradigmatic case of confined two-dimensional lyotropic systems, the formation of {\rene grain boundaries largely} dominates over strong elastic deformations \cite{slitpores,CollLQinSqConf,annulus}.
{\rene At these grain boundaries, the nematic order present in the bulk smectic phase breaks down,
hindering a classification of the emerging orientational patterns in terms of nematic topology alone.}

\begin{figure}[t]
\begin{center}
\includegraphics[width=0.9\linewidth]{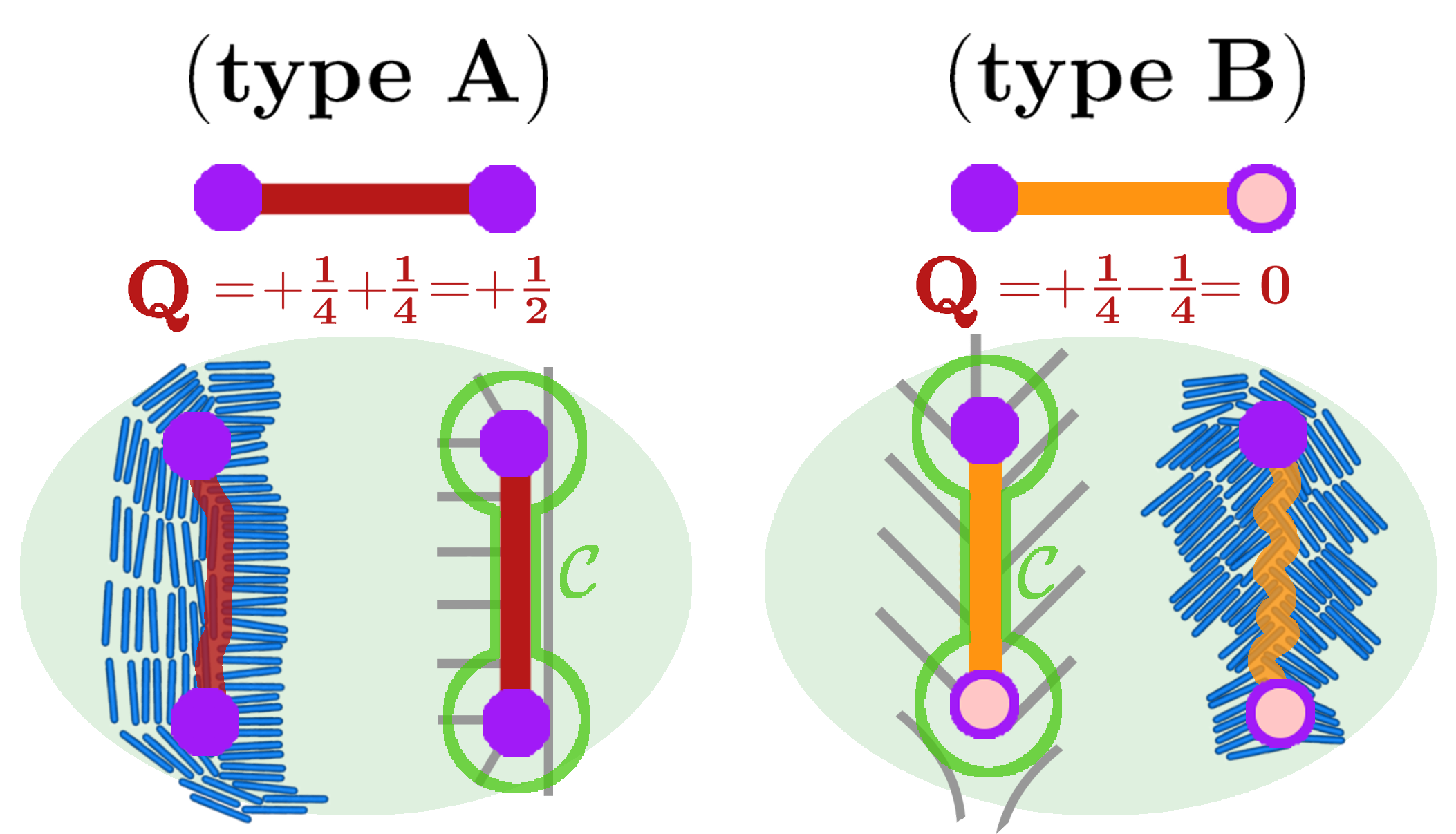}
\includegraphics[width=0.8
\linewidth]{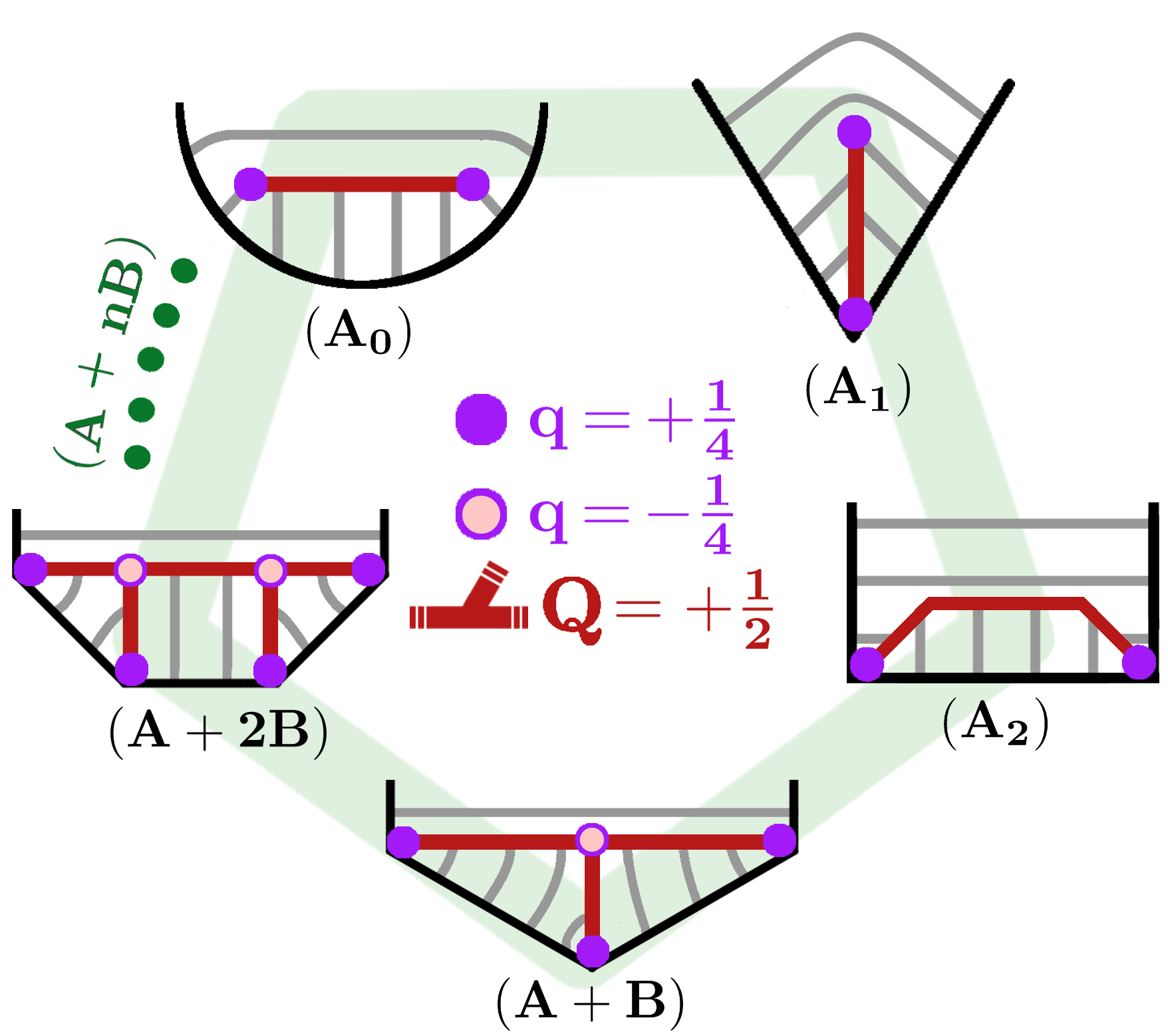}
\caption{Survey of {\rene grain boundaries with different connectivities}
in smectic liquid crystals {\rene at a convex confining wall}.
{\rene \textbf{Top:} composition of the} two fundamental building blocks with total {\rene nematic} charge $\Qnem=1/2$ ({type A}, {\rene red line}) and $\Qnem=0$ ({type B}, {\rene orange line}),
{\rene determined by the tetratic quarter charges $\Qtet$ (dots) at the end points}.
{\rene The illustration of bulk} orientational ordering 
 depicts a typical arrangement of the surrounding rods 
and a schematic continuum picture with idealized straight {\rene grain boundaries} separating regions of perpendicular smectic layers (grey lines). 
{\rene The closed contour $\mathcal{C}$ (green line) highlights the contribution of the tetratic end points to $\Qnem=\sum\Qtet$.
\textbf{Bottom:} relation} between the geometry-dependent manifestations of $\Qnem=+1/2$ {\rene grain-boundary networks (red) in smectics confined to polygons}. 
The simplest structure $(A_m)$ only contains one type-$A$ defect with $m\in\{0,1,2\}$ of its end points attached to a corner of the {\rene confining wall}.
In general, complex networks $(A\!+\!nB)$ can form, which amounts to adding $n$ building blocks of type $B$. 
Approaching the limit $n\rightarrow\infty$, the network detaches from the increasingly smooth {\rene corners, gradually reverting to $(A_0)$, as} the {\rene tetratic defects in the} type-$B$ branches annihilate {\rene \cite{SI}.}
}
\label{fig_BuildingBlocks}
\end{center}
\end{figure}

{\rene 
In this Letter, we demonstrate that 
extremely confined smectic systems can be effectively described in terms of topological defects in the tetratic order due to the strong preference of smectic layers to tilt at a grain boundary by approximately $90^\circ$.
 The tetratic topology is thus not only important in 
systems with tetratic bulk symmetry \cite{li2014topological, narayan2006nonequilibrium,granular0,armas2020} 
but also a vital ingredient to a comprehensive picture of frustrated smectics.
In detail, we identify quarter-integer tetratic {\reneN disclinations}, 
which materialize in pairs at the extremities of {\rene grain boundaries},
as the elementary topological unit of smectic liquid crystals.}
In turn, the notion of a nematic {\reneN disclination} 
expands to a spatially extended defect structure
whose half-integer charge follows
from the sum of its tetratic components, {\rene thereby acting as a spatial charge distribution,} as exemplified in \figref{fig_Lupenbild}.
To {\rene unveil the full implications  of coexisting nematic and tetratic order}, we use particle-resolved computer simulations and colloidal experiments on hard rods to create
different defect structures in a large range of two-dimensional geometries.
{\rene Defining the grain boundaries connecting different types of tetratic \reneN disclinations} 
as fundamental building blocks, we {\rene provide the basic toolbox to} characterize the more complex smectic defect networks 
emerging in the presence of multiple corners, as illustrated in \figref{fig_BuildingBlocks}. 
Thereby, our approach visualizes the topological charge conservation {\rene $\sum \Qtet=\sum \Qnem=1$} for {\rene both individual tetratic point defects $\Qtet$ and}
  defect networks {\rene with nematic charge $\Qnem$},
whose typical connectivity depends on the {\rene curvature landscape of the confining wall}. 

To create the smectic structures for each confinement, we perform canonical Monte Carlo (MC) simulations on $N=1000$ hard discorectangles \cite{overl} of aspect ratio $p=15$ within two-dimensional cavities, bounded by WCA-like soft walls \cite{WCA,metropolis}, see Supplemental Material \cite{SI}. We randomly initialize the system at a low area fraction $\eta_0 \ll 1$. The system is quickly compressed at a rate of $\Delta \eta_1 = 4.15 \times 10^{-7}$ per MC cycle to an area fraction $\eta_1 \approx 0.29$, where the isotropic-nematic transition is expected \cite{phase_beh_DF}. We subsequently compress the system at a lower rate $\Delta \eta_2 = 4.625 \times 10^{-8}$ per MC cycle
until the system reaches the target area fraction of $\eta_2 = 0.75$ and exhibits smectic order.
This protocol ensures that the system is close to equilibrium at all times.
After equilibration, we use cluster analysis to identify domains with different orientational order and generate statistics. For each state, we determine
 local {\rene nematic $S (\mathbf{r})$ and tetratic $T (\mathbf{r})$ order parameter fields} to identify the {\rene composition of} topological defects \cite{SI}.

\begin{figure*}[t]
\begin{center}
\includegraphics[width=1.00\linewidth]{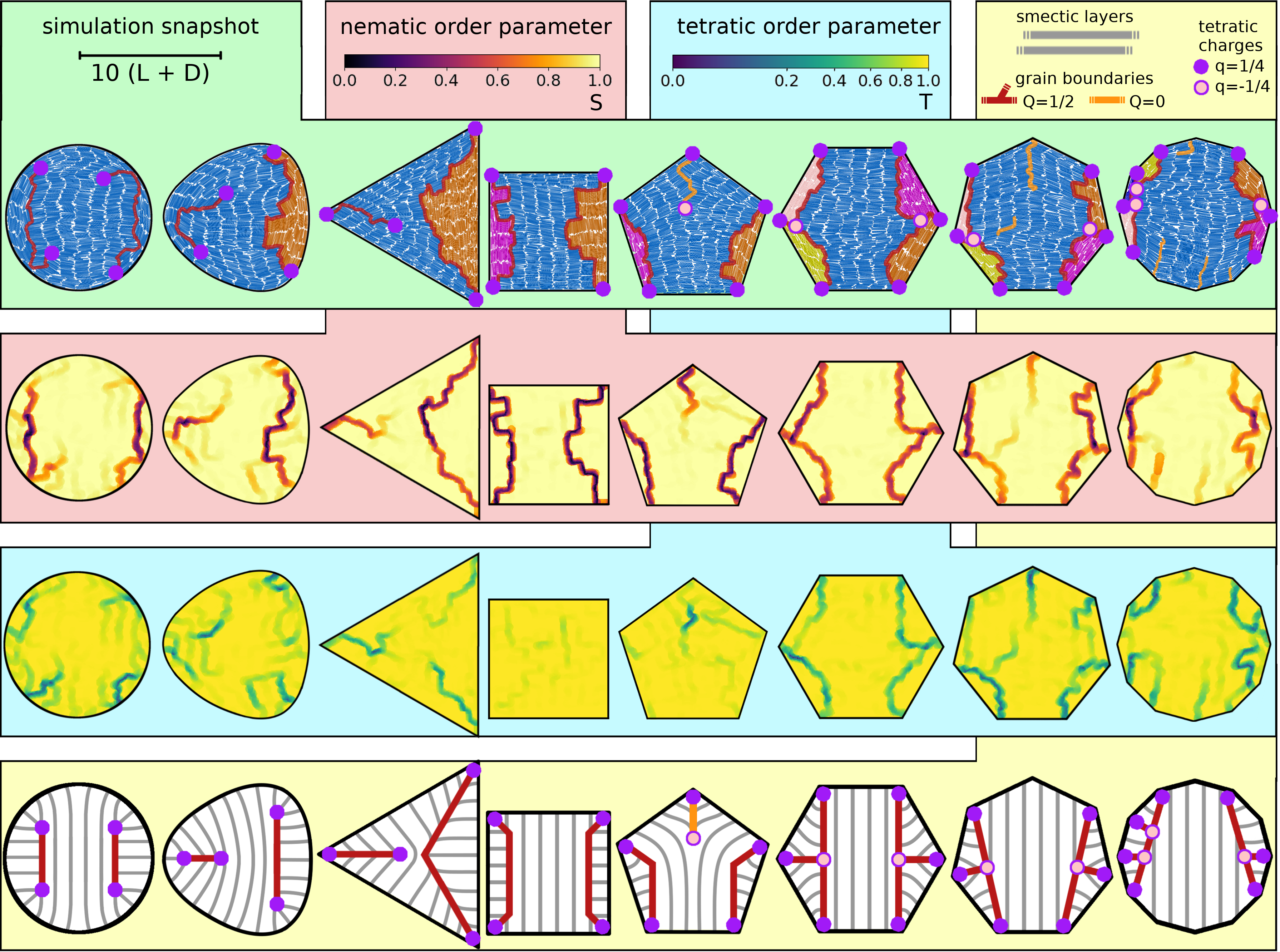}
\caption{
Topological defect structure of representative simulations for hard rods with aspect ratio $p=15$ in different convex confining geometries.
\textbf{Top row:} particle snapshots with superimposed networks {\rene of grain boundaries} {\rene and isolated tetratic point defects}, compare Fig.~\ref{fig_BuildingBlocks}.
{\rene The color of the rods 
highlights individual domains according to cluster analysis.}
\textbf{Second row:}
{\rene nematic} order parameter field \Sofr. 
\textbf{Third row:}
{\rene tetratic order parameter field $T\ofr$. 
Point defects at the confining wall are not visible}.
\textbf{Bottom row:} idealized continuum interpretation of the depicted snapshots,  {\rene as detailed in the Supplemental Material \cite{SI}.} 
}
\label{fig_raster}
\end{center}
\end{figure*}

On the experimental side, we analyze smectic structures emerging at the bottom of tailored cavities at sedimentation-diffusion equilibrium of colloidal silica rods \cite{CollLQinSqConf}.
The synthetic rods \cite{kuijk2012} have a small polydispersity in length and diameter.They are dispersed into a $1\,$mM NaCl aqueous solution \cite{annulus},
which leads to a short-ranged repulsion and effective hard-rod-like interactions.
The degenerate planar anchoring at the bottom wall allows us to capture images of quasi-two-dimensional smectic states in the horizontal plane.
Using bright-field microscopy with an objective of high numerical aperture, we can discriminate most rods and thus determine the local order.

Our numerical observations are summarized in \figref{fig_raster}.
The common feature of all structures is a large, defect-free central domain,
characteristic for the bridge state \cite{slitpores,CollLQinSqConf,annulus}.
The detailed appearance of the topological defects, however, sensitively depends on the confining geometry.
To verify that the overall topology is universal, 
we decorate {\rene \cite{SI}} all representative snapshots with a defect structure assembled from the building blocks in \figref{fig_BuildingBlocks}.
 By doing so, we recognize in each system in \figref{fig_raster} two separate {\rene grain-boundary} networks representing a $\Qnem=+1/2$ charge each, while the remaining defects do not carry any {\rene nematic charge $\Qnem$} as a whole.
The intriguing dependence of the emerging defect networks
on the geometric properties of the confining {\rene wall} can be perceived according to the schematic cycle in \figref{fig_BuildingBlocks},
{\rene as laid out in the following three paragraphs}.

The {\rene most common defects are linear grain boundaries} of the general type $A$,
which we further discriminate by the location of the two {\rene positively charged tetratic} end points, cf.~\figref{fig_BuildingBlocks}.
The circular cavity in \figref{fig_raster} {\rene typically} features two opposing 
bulk defects of subtype $A_0$, i.e., both end points {\rene possessing isolated tetratic signals} are detached from the {\rene wall}.
Therefore, the orientation of rods around the perimeter changes continuously and all particles within the system belong to a single domain. 
Upon switching from the uniformly curved circular confinement to different polygons, 
the {\rene grain boundaries usually extend towards} the corners, 
{\rene such that the tetratic defects anchor at the wall. 
The invariance of our topological picture 
can be illustrated by considering confinements with smooth 
corners \cite{SI}.
The example of a rounded equilateral triangle} in \figref{fig_raster} indicates that one {\rene grain boundary} turns into a {\rene true} domain boundary, i.e., a type-$A_2$ defect, as both its end points move towards two of the three corners.
The {\rene second grain boundary} gradually turns {\rene into a type-$A_1$ defect}, as one end point attaches with the remaining corner and the other end extends into the center of the large domain.
{\rene This structure is} most pronounced in the limit of sharp corners.
{\rene We} further observe 
{\rene that a grain boundary of type $A_1$ gradually}
contracts to a $\Qnem=+1/2$ point defect upon decreasing the opening angle \cite{SI}.
Turning to a square cavity in \figref{fig_raster}, the additional corner can accommodate the loose end point of the type-$A_1$ defect, {\rene resulting in} the eponymous bridge structure \cite{CollLQinSqConf,slitpores} with two {\rene parallel} type-$A_2$ domain boundaries, {\rene as in circular confinement, but with perfect tetratic bulk order}.

\begin{figure}[t!]
\begin{center}
\includegraphics[width=1.0\linewidth]{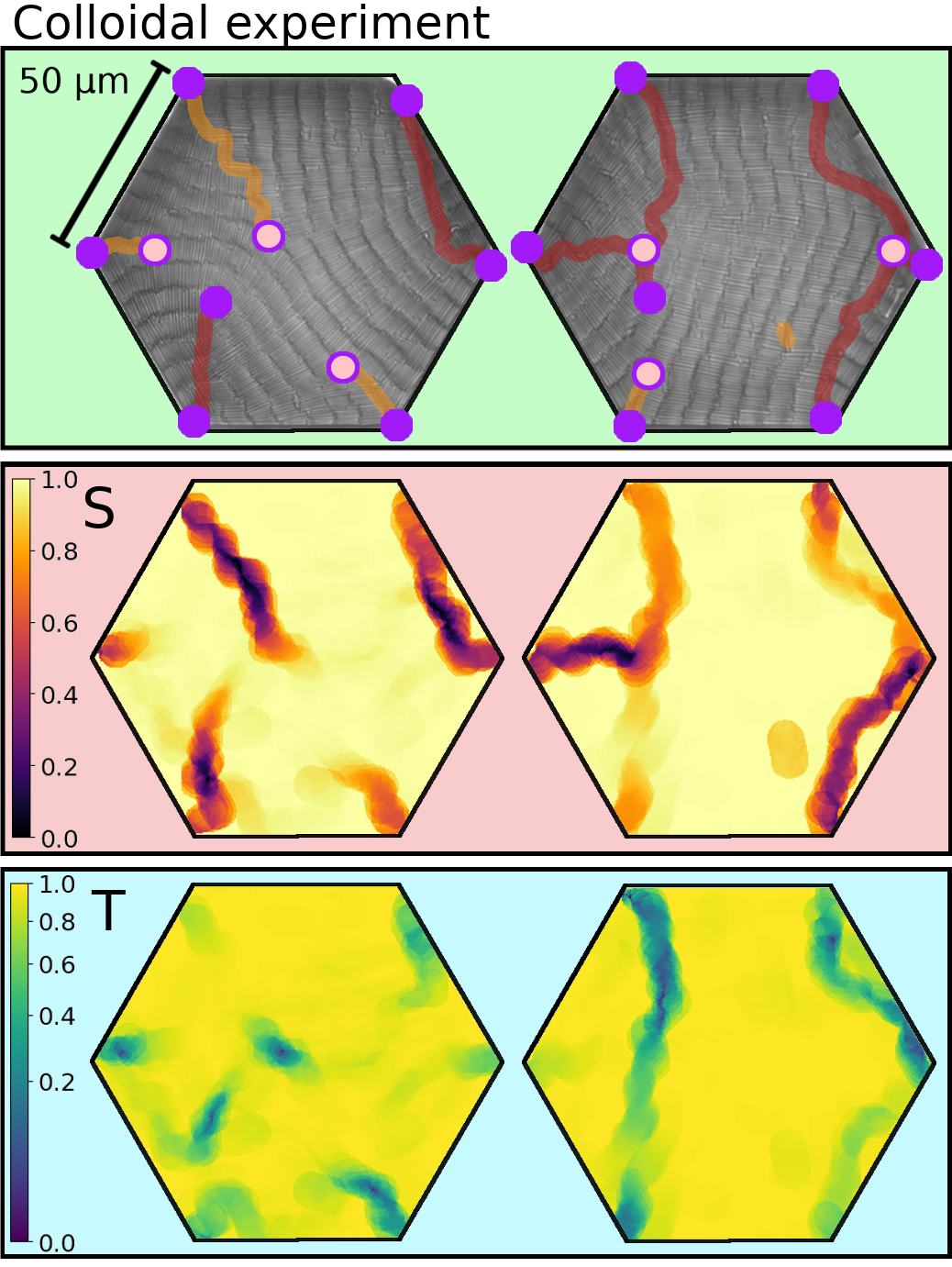}
\caption{Two selected sets of experimental reference data in hexagonal confinement, {\rene presented as in Fig.~\ref{fig_raster}.
Shown are bare bright-field microscopy images with $N=1400\pm150$ colloidal rods of effective hard-rod aspect ratio $p_\text{eff}=10.6$ in the field of view and the extracted order parameters.}
}
\label{fig_experiment}
\end{center}
\end{figure}

{\rene Following in} \figref{fig_raster} a sequence of geometries represented by regular polygons,
the increasing number of {\rene $\Qtet=+1/4$ charges at each corner is} compensated accordingly by negative {\rene bulk} charges.
{\rene We thus introduce an additional type-$B$ building block representing a pair of tetratic $\Qtet=\pm1/4$ charges. This overall charge-neutral object with $\Qnem=0$} 
either occurs on its own
 {\rene (at edge dislocations in the bulk or attached to a single corner)}
or attaches with its {\rene negative} end to other building blocks, forming a large network of {\rene grain boundaries}, cf.~\figref{fig_BuildingBlocks}. 
 {\rene In particular, the pentagon can accommodate
 an individual type-$B$ defect in addition to the two type-$A_2$ domain boundaries
also found in square confinement}
(notice the difference with the free-standing type-$A_1$ {\rene defect with $\Qnem=1/2$} in the triangle).
 {\rene The hexagon typically} features {\rene two parallel type-$(A\!+\!B)$} networks, each separating two small domains from the central bridging layers 
 {\rene and resembling a}
type-$A$ defect, like in the square, with an attached
type-$B$ defect emerging from the additional corner in the middle.

{\rene The addition of further corners allows for the formation of} complex type-$(A\!+\!nB)$ networks, which contain $n$ nodes {\rene with $\Qtet=-1/4$ and $n+2$ branches ending on a $\Qtet=+1/4$ charge.
The typical defect structure, however, gradually}
reverts to that in circular confinement, {\rene closing the cycle in \figref{fig_BuildingBlocks}}.
The defect networks in \figref{fig_raster} thus detach from the confining wall, as adjacent pairs of opposite tetratic charges annihilate.
{\rene Grain boundaries between pairs of defects close to annihilation typically
induce only a small tilt between smectic layers,
such that their nematic signal weakens and qualitatively resembles the tetratic signal
(contrast, e.g., the free-standing type-$B$ defects in the pentagon and heptagon).
\rene In general, the degree of the annihilation increases with increasing opening angle at the corners~\cite{SI}.}

To demonstrate the experimental relevance of our {\rene classification scheme, we analyze in} Fig.~\ref{fig_experiment} microscopy images of colloidal rods for their orientational order, here focusing on a hexagonal domain.
{\rene The experimental defect networks are typically less complex than those in the pure hard-rod simulations \cite{SI}, due to the higher elasticity of the smectic layers. Nonetheless, the defect networks found consist of the same fundamental building blocks, confirming the broad applicability of our approach. Additionally, our simulation results match the previously reported experiments in square \cite{CollLQinSqConf} and circular \cite{annulus} confinement, and hence the same analysis is directly suitable to those experiments as well.}

Beyond the chosen methodology, our {\rene topological toolbox}
can be readily {\reneN employed to illustrate the coexisting nematic and tetratic orientational order of frustrated smectics in free-energy based theoretical studies \cite{PD_Rene,annulus,wittmann2016,pevnyi2014,jingmin2021},
granular experiments \cite{narayan2006nonequilibrium,granular0,armas2020} or molecular systems~\cite{originalPointsToLines,michel2006moleculardefect,coursault2016moleculardefect}.
Regarding nonconvex confinements \cite{annulus} or two-dimensional manifolds \cite{mosna2012,allahyarov2017}, 
our set of building blocks can be supplemented by a line connecting two negative tetratic charges. 
A generalized approach can shed more light on intersecting surfaces of grain boundaries in three-dimensions.}
{\rene One central implication of our 
analysis is that the motion of grain boundaries can be tracked through tetratic point defects,
\reneN providing additional insights into}
 the dynamics of smectics \cite{dhont1996introduction,grelet2008dynamical,kurita_tanaka2017,chiappini2020}
and their nucleation \cite{schilling2004,ni2010,cuetos2010},
{\reneN which is particularly interesting for biologically inspired} nonequilibrium systems like self-propelled rods \cite{REVbaerSPR2020,kumar2019,maitra2020,bott2018} 
or growing bacterial colonies \cite{bacterialcolonies2020a,bacterialcolonies2020c,bacterialcolonies2021} {\rene as candidates for active smectics}.

{\rene  Finally, we expect that the classification of the fine structure of defects  on the length scale of individual particles put forward in this work 
will be helpful to analyze quenched or undercooled systems, in particular those where the symmetry of an ordered phase is broken by grain boundaries that may still impose a preferred alignment between adjacent domains. One possible example would be fine-grained polycrystals, which are challenging to distinguish from amorphous solids on the single-particle level \cite{zhang2018compression}.
More generally, those methods could lead to a better understanding of defects in complex solids
(such as protein \cite{tsekova2012effect} or aerosol \cite{dusek2006size} crystals)
 relevant for photonics \cite{ling2014photonic}, phononics \cite{guilian2016phononic} and metamaterials \cite{wang2020metamaterials}. }

The authors would like to thank Daniel de las Heras, Axel Voigt, Raphael Wittkowski and Michael te Vrugt for helpful discussions.
This work was supported by the German Research Foundation (DFG) within project LO 418/20-2.

P.M. and R.W. contributed equally to this work.

\newpage\mbox{}
\newpage

\begin{widetext}

\setcounter{figure}{0}
\setcounter{equation}{0}
\renewcommand\theequation{S.\arabic{equation}}

\noindent
\begin{center}
{\Large\bf\centering Supplemental Material}
 \end{center}\bigskip

\section{Numerical methods}

\subsection{Model system}

\label{sec_model}
We model our system as fluid of hard discorectangles that are defined as rectangles with length $L$ and width $D$ capped by two half discs with diameter $D$, interacting via pure excluded-volume interaction \cite{overl}. 
We focus on particles with length-to-width ratio $p=L/D = 15$. 
The coordinates of the particles are defined via their position vectors $\mathbf{r}_n$ and orientation vectors $\mathbf{\hat{u}}_n=(\cos\phi_n,\sin\phi_n)^T$, where $\phi_n$ is the angle of the $n$th particle with an arbitrary fixed axis.
The particles are confined within various two dimensional cavities as depicted in Fig.~3 in the main manuscript.
To force particles from the outside of the cavity to the inside, we model the interaction of the particles with the walls as a soft cut-off  Weeks-Chandler-Anderson (WCA) potential \cite{WCA}. 
To determine the interaction energy of the particles with the walls in our model, we consider the discoidal caps sitting at $\mathbf{r}_n \pm L/2~\mathbf{\hat{u}_n}$.
Defining  the relative positions $x_+$ and $x_-$ of both discs to the walls of the confinement, we compute the wall energy of the $n$th rod as 
\begin{align}
V(\mathbf{r}_n,\mathbf{\hat{u}_n}) = V(x_+) + V(x_-)
\end{align}
with the potential  
\begin{align}
\label{eq_wallpot}
    V(x) =
\begin{cases}
 \Phi(x_{0}) +\Phi'(x_{0})(x-x_0)  & \text{ for }  x \leq x_{0}  \\
 \Phi(x)  & \text{ for }  x_{0} > x  
\end{cases}
\end{align}
providing a version, linearized below $x_0 = D/2$ and thus extended to the outside of the cavity, of the regular WCA-potential
\begin{align}
\label{eq_WCA}
    \Phi(x) =
\begin{cases}
4\epsilon \left [  \left ( \frac{D}{x} \right )^{12} - \left( \frac{D}{x} \right )^{6} \right ] + \epsilon
& \text{ for }  x/D \leq 2^{\frac{1}{6}}
\\
 0  & \text{ for }  x/D >2^{\frac{1}{6}}
\end{cases}\,,
\end{align}

where we use the energy scale $\epsilon = 10$ $ k_\text{B} T$ to mimic nearly hard walls,
thereby imposing a bias for planar surface anchoring. 

\subsection{Simulation procedure} \label{sec_simproc}
We study systems at constant particle number $N$ and constant temperature $T$ using standard canonical Monte-Carlo simulations. 
We define the area fraction of the confined particles as $\eta = {N A_\text{HDR}}/{\mathcal{A}}$ with the area of a hard discorectangle
\begin{equation}
    A_\text{HDR}= LD + \frac{\pi D^2}{4}
\end{equation}
and the area $\mathcal{A}$ of the confining geometry, assuming hard walls at $x=0$.
We focus on the study of smectic states at an area fraction $\eta \approx 0.75$, where the smectic phase is stable in bulk \cite{PD_Rene}.

Every simulation runs for at least $10^7$ Monte-Carlo cycles. One cycle consists of $N$ trial moves where we randomly choose one particle and, by equal probability, either displace it or rotate it. The particles are displaced by changing their $x$- and $y$-coordinates by $\Delta_{x,y} \in [-\Delta_{x,y}^{\text{max}},\Delta_{x,y}^{\text{max}}]$. Rotation is done by displacing the orientation vector $\mathbf{\hat{u}}_n$ along its orthogonal by $\Delta_{\mathbf{\hat{u}}} \in [-\Delta_{\mathbf{\hat{u}}}^{\text{max}},\Delta_{\mathbf{\hat{u}}}^{\text{max}}]$ and then renormalizing.
The whole trial step is accepted with the standard acceptance probability $P_{\text{acc}} = \min(e^{\beta \Delta U},1)$ \cite{metropolis} where $\Delta U$ is the total change in the energy of the system.
The rate with which particle trial moves are accepted depends on  $\Delta_{x,y}^{\text{max}}$ and $\Delta_{\mathbf{\hat{u}}}^{\text{max}}$. Those quantities are adjusted over the course of the simulation to stabilize the acceptance rate at about $0.1$.
 
To obtain the desired high-density states, we initialize our simulation by randomly placing all $N$ particles within an augmented confinement, such that the initial area fraction is $\eta_0 \ll 1$. We then follow a compression protocol to bring the system slowly to the desired density.
This is done at sufficiently low compression rates to allow the system to reach thermodynamic equilibrium at each density.
Specifically, the compression procedure is divided into two stages: a first one 
where we increase the area fraction of by shrinking the confinement with a relatively high compression rate $\Delta \eta_1 = 4.15 \times 10^{-7}$ per MC-cycle to an area fraction $\eta_1 \approx 0.29$
where nematic ordering starts to occur. 
This is observed to lie slightly below the area fraction of the isotropic to nematic phase transition \cite{phase_beh_DF}. Once the system reaches $\eta_1$ we reduce the compression rate to $\Delta \eta_2 = 4.625 \times 10^{-8}$ per MC cycle and compress further until the desired target area fraction $\eta_2$ (we typically choose $\eta_2 = 0.75$) is reached and we sample the results.
As there is a significant difference between nematic and smectic states in a square cavity, we equilibrated the systems confined to square cavities at compression rates that were slower than those for the other geometries by a factor of four.

\subsection{Data analysis}
\label{sec_analysis}

In this section, we define the local quantities, we use characterize the structure of the confined systems.
The central {\rene quantities to our topological analysis are the nematic   
\begin{align}
\label{eq_Sofr}
    S \ofr = \Betrag{ \langle e^{i 2 \phi_n} \rangle_\mathbf{r}}\,
\end{align}
and the tetratic
\begin{align}
\label{eq_Tofr}
    T \ofr = \Betrag{ \langle e^{i 4 \phi_n} \rangle_\mathbf{r}}\,
\end{align}
orientational order parameters,
where the angled brackets $\left \langle ... \right \rangle_\mathbf{r}$ denote the average over all particles which intersect a local circle around $\mathbf{r}$ with radius $\xi = 5D$. 
\Sofr~and $T \ofr$ take values in the interval $[0,1]$, where $0$ denotes low local orientational order and $1$ corresponds to high local orientational order.
Furthermore we measure the local nematic director angle $\psi_2\ofr$ and tetratic director angle  $\psi_4\ofr$ as
\begin{align}
     \psi_m\ofr = \frac{1}{m}  \arccos \left ( \Re \left (  \frac{\langle e^{i m \phi_n}\rangle_\mathbf{r}}{\Betrag{ \langle e^{i m \phi_n} \rangle_\mathbf{r}}} \right ) \right ) 
\end{align}
with $m\in\{2,4\}$ and the real part $\Re(...)$ of a complex number.
The normalized nematic \nofr~and tetratic director fields \tetr$(\mathbf{r})$ are defined as \nofr$=\cos(\psi_2) \mathbf{\hat{e}_x} + \sin(\psi_2) \mathbf{\hat{e}_y}$
and \tetr$(\mathbf{r})=\cos(\psi_4) \mathbf{\hat{e}_x} + \sin(\psi_4) \mathbf{\hat{e}_y}$, respectively.}

To {\rene facilitate the identification of grain boundaries and to} determine the number of the domains within the confinement, we perform a cluster analysis on the final equilibrated configuration based on the inter-particle distances and the relative orientations. 
We chose the pairwise domain criterion such that two particles $k$ and $l$ are assigned to the same domain if $(i)$ the distance of their position vectors is smaller than a threshold $\Betrag{\mathbf{r}_k-\mathbf{r}_l} \leq 1.3~(L+D)$ and $(ii)$ their relative orientation follows the criterion
\begin{equation}
\label{eq_clustercritB}
\left | \mathbf{\hat{u}}_k \cdot \mathbf{\hat{u}}_l \right | \geq 0.95\,.
\end{equation}

{\rene To further identify the fine structure of the system, we draw bonds between neighboring particles that we consider to be in the same smectic layer.
For each particle $i$ with position $\mathbf{r}_i$ and orientation $\mathbf{u}_i$, 
we also define the vector  $\mathbf{u^\perp}_i$, which is taken to be perpendicular to $\mathbf{u}_i$.
We then consider two possible neighbors: the particle that is closest to the point $\mathbf{r}_i + \mathbf{u^\perp}_i$ and the particle that is closest to $\mathbf{r}_i - \mathbf{u^\perp}_i$. By closest, we here mean the distance of the chosen point to any point on the particle's surface.
A chosen particle $j$ on each side of particle $i$ is now considered a neighbor if
\begin{enumerate}
    \item $|\mathbf{\hat{u}}_i \cdot \mathbf{\hat{u}}_j| > 0.9$, and
    \item $|\mathbf{r}_{ij} \cdot \mathbf{\hat{u}}_i| < 0.45 L$, and
    \item $|\mathbf{r}_{ij} \cdot \mathbf{\hat{u}}_j| < 0.45 L$,
\end{enumerate}
where $\mathbf{r}_{ij} = \mathbf{r}_i - \mathbf{r_j}$.
To ensure symmetry, we remove any bonds from particle $i$ to $j$ where particle $j$ is not bonded to particle $i$.
}

\section{Characterizing the topological defect structure}%
\label{sec_top_def_struc}
\begin{figure}[t]
 \centering
{\includegraphics[scale=0.4]{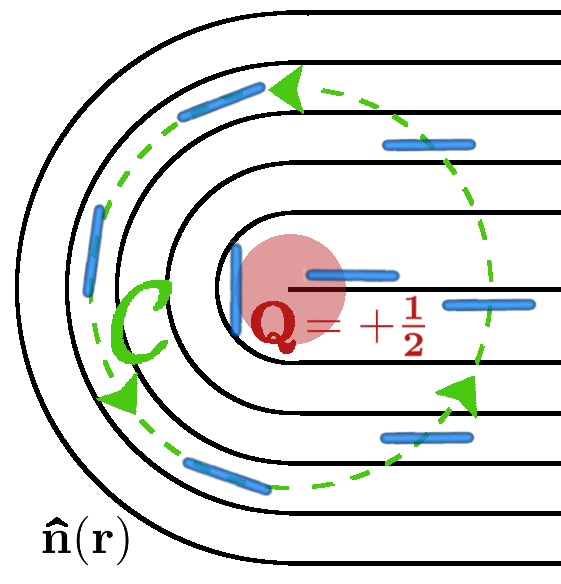}}
\caption{
Schematic depictions of the topological charge of a singularity in the director field \nofr, illustrated for a pointlike $\Qnem=+1/2$ {\reneN disclination}. The topological charge is calculated according to Eq. (\ref{eq_Q}) via the winding number along a closed contour $\mathcal{C}$ (green). Circling around the defect once, the particles orientation changes by an angle of $\pi$, which results in $q = +1/2$. 
}

\label{fig_schem_defects}
\end{figure}

\subsection{Topological charge conservation of pointlike {\rene nematic} bulk and boundary defects \label{sec_topN}}

{\rene To introduce the fundamental topological concepts relevant to determine and interpret the charge of a defect,
we first reiterate the established methods for a fluid with pure nematic symmetry. 
Nematic} orientational order of uniaxial liquid crystals is described by 
the director field \nofr, with the identification $\mathbf{\hat{n}} \equiv -\mathbf{\hat{n}}$, reflecting the apolarity of the particles. 
To determine the charge of the topological defects emerging in a two-dimensional system, 
we consider a closed contour $\mathcal{C}$ with one revolution in counter-clockwise direction, as illustrated in \figref{fig_schem_defects},
assuming that $\mathcal{C}$ does not pass any points or regions with a discontinuous {\nofr}.   
Hence, the local orientation of the director will always return to its initial value,
when traversing once along a given $\mathcal{C}$ from any starting point.
This results in the topological charge $\Qnem$, defined as the number of revolutions of the director relative to those of $\mathcal{C}$, being always a multiple of $1/2$, see \figref{fig_schem_defects}a for an example with $\Qnem=1/2$. 
Introducing a parametrization $\kappa$ of the contour $\mathcal{C}(\kappa)$, the topological charge
can be explicitly calculated according to \cite{SoftMPhys,lavrentovich2010topological} 
\begin{align}
\label{eq_Q}
\Qnem = \frac{1}{2\pi} \oint_{\mathcal{C}(\kappa)}~\left [\mathbf{\hat{n}} (\kappa) \times \frac{\partial \mathbf{\hat{n}} (\kappa)}{\partial \kappa} \right ] d\kappa\,,
\end{align}
where $\oint_{\mathcal{C}(\kappa)} d\kappa=2\pi$.
Since $\Qnem$ takes only discrete values but must change continuously whilst continuously altering the path $\mathcal{C}$ in a region without singularities, it has to be constant in such a region \cite{algebTop}. 
Moreover, the total charge within a given domain can be computed as the sum of individual charges.

\begin{figure}[t!!!!!!!!!]
\begin{center}
\includegraphics[width=0.8809\linewidth]{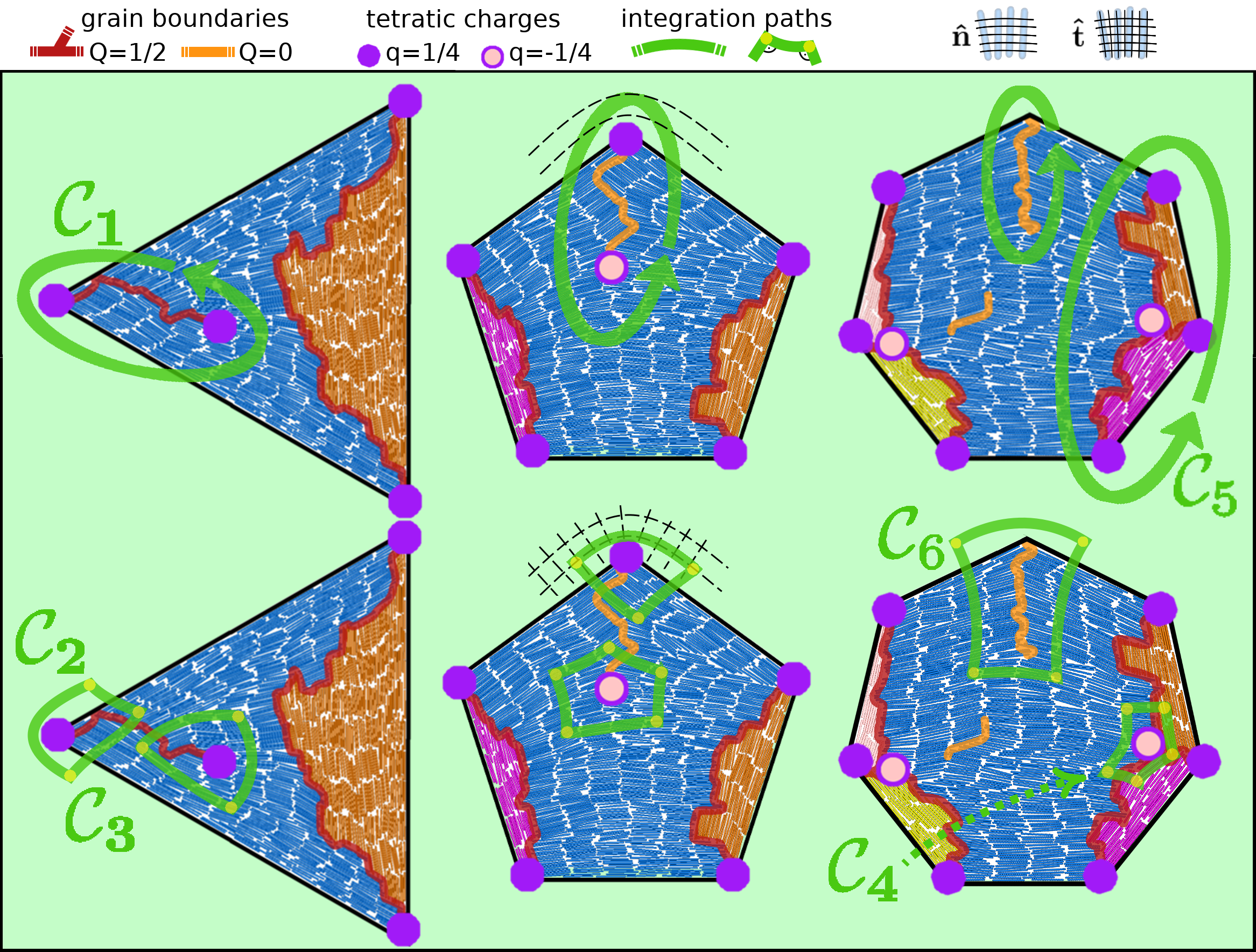}
\caption{
\rene Overview of how to determine the topological charge distribution (superimposed points and lines) within different simulation snapshots. 
Each defect represents a discontinuity in the director field,
 which can be enclosed by a closed contour $\mathcal{C}$ (green paths),
only traversing regions with a continuous director. 
Virtually extending the director fields to the region outside the cavity (representative black lines) allows to assign a topological bulk charge to each defect, 
irrespective of its connectivity with the system boundaries.
The labeled contours are explicitly referred to in the text.
\textbf{Top row:} The nematic director field~\nofr is parallel to the rod axes and thus discontinuous at a 
grain boundary.
Every network of grain boundaries, i.e., an assembly of interconnected nematic line defects 
can be assigned a charge $\Qnem$ as a whole by integrating up the local rotation angle of \nofr~along $\mathcal{C}$, according to Eq.~(\ref{eq_Q}). 
\textbf{Bottom row:} The field lines of the tetratic director \tetr$\ofr$ are both parallel and perpendicular to the rod axes 
and thus continuous at a grain boundary with typical deficiency angle $90^{\circ}$, while the rotation of the field happens around the end points.
The charge $\Qtet=1-c/4$ of these tetratic defects follows from Eq.~(\ref{eq_q}) 
and can be directly inferred from counting the number $c$ of $90^{\circ}$-corners (yellow points)
of a polygonal integration loop $\mathcal{C}$ on top of the field lines.
A grain-boundary network can therefore be understood as a consolidation 
of isolated tetratic defects.
As one principal axis of \tetr$\ofr$~is always equal to \nofr, the nematic charge of such a network 
always equals the total tetratic charge of its end points and nodes according to Eq.~(\ref{eq_conservationTET}), compare contour $\mathcal{C}_1$ with $\mathcal{C}_2$ and $\mathcal{C}_3$.
Grain boundaries with a deficiency angle $<90^{\circ}$ can only be assigned a topological charge $\Qtet=\Qnem$ as a whole, see contour $\mathcal{C}_6$.
}

\label{fig_new_contours}
\end{center}
\end{figure}

In a confined system, the external boundary induces a particular alignment of the particles,
which is preferably parallel to the walls in case of hard rods at a hard wall.
This typically enforces the formation of defects.
{\reneN If all particles obey a strong tangential anchoring condition at each point of the boundary, then the} sum of all topological charges is determined by the Euler characteristic $\chi=1$ of the simply-connected confining domains considered here \cite{algebTop}.
The corresponding fundamental law of charge conservation reads
\begin{align}
    \sum_i \Qnem_i=1\,, 
    \label{eq_conservationlaw}
\end{align}
where the index $i$ labels all defects in the system.

{\reneN The topological defects occurring in nematic liquid crystals can be generally classified according to their location relative to the confining walls as bulk, boundary or virtual defects \cite{VirNemConfGeom,garlea2016finite}.
The charge of a bulk defect determined according to Eq.~\eqref{eq_Q} is not affected by the presence of the boundary.
A boundary defect indicates a violation of the anchoring condition.
This is particularly the case at a corner, i.e., a singular point of the boundary curvature,
which is not compatible with the symmetry of the director field \nofr.
The sudden jump of  \nofr~when traversing the confining wall can be related to the deficiency angle $\tau=180^\circ-\alpha$ at the corner of opening angle $\alpha$, which leads to the common definition \cite{real_defch,lavrentovich2010topological,VirNemConfGeom}
$m = k/2 - \tau / (2\pi)$ of a of a boundary charge $m$, where $k$ takes integer values.}
The corresponding conservation law generalizing Eq.~\eqref{eq_conservationlaw} then reads 
\begin{align}
    \sum_i \left ( m_i + \frac{\tau_i}{2\pi} \right ) + \sum_j \Qnem_j = 1\,,
    \label{eq_conservationGEN}
\end{align}
with the bulk defect charges $\Qnem_j$.
Finally, a virtual defect represents a director field which does not uniformly align with the boundary, 
such that the center of this distortion can be thought to lie outside the actual system \cite{garlea2016finite}.
For smectic liquid crystals, which also exhibit positional order, 
the interpretation of the defect structure, described in the remainder of this section, requires further care.

\subsection{ {\rene Nematic and tetratic} topological charge within smectic  {\rene grain boundaries} \label{sec_topSM}}

 In this section we describe the classification of topological defects in a smectic fluid, which are of central interest in our simulations and experiments.
Regarding the spatially extended {\rene defect structure represented by the} emerging smectic {\rene grain boundaries},
a clear distinction between bulk and boundary defects is not always possible. 
For example, depending on the local curvature of the wall, we observe a continuous transition between {\rene grain-boundary} lines representing pure bulk defects 
and those with one or both end points attached to the boundary  (see, e.g., Fig.~\ref{fig_rounded_tria_raster}).
{\rene On a second level of our classification, the end points of grain boundaries themselves can be identified as tetratic bulk or boundary defects,
which could be classified along the lines of Sec.~\ref{sec_topN}. However,}
in what follows, we {\rene seek for} a unifying {\rene virtual} treatment for all {\rene defects, irrespective of their location, which we illustrate for both nematic and tetratic order in Fig.~\ref{fig_new_contours}}. 

{\rene
As a first step, we introduce a virtual extension of the director field that is tangential to the boundary and continuously circulates each corner in the outside region. 
{\rene Hence, we avoid the ambiguity arising from the angular deficiency at the corners and treat all defects as bulk defects}, {\rene compare, e.g., the contour $\mathcal{C}_1$ in Fig.~\ref{fig_new_contours}}. 
This picture allows us to associate a well-defined topological charge $\Qnem \in\{\pm 1/2,\pm 1, \pm 3/2~\ldots\}$, determined via Eq.~\eqref{eq_Q}, with any {\rene defect in the nematic order},
irrespective of its spatial extension and connectivity with the boundary, {\rene see Fig.~\ref{fig_new_contours} for representative examples}.
A pointlike defect at the system boundary, in particular, is thus interpreted as a virtual defect with half-integer topological charge  $\Qnem=m + \tau / (2\pi)$, such that the charge conservation law for internal bulk charges from Eq.~\eqref{eq_conservationlaw}, {\reneN which applies to our classification,} is recovered from Eq.~\eqref{eq_conservationGEN}.
{\rene This concept is best justified regarding the detachment of defects from the boundary in polygons with rounded corners, as detailed in Sec.~\ref{sec_rtria}.}

As a second step, we apply the continuum assumption
of a tilt angle between smectic layers at opposite sides of grain boundaries
that is exactly $90^{\circ}$.
Therefore, each grain boundary is compatible with tetratic orientational order, described by the director field \tetr$\ofr$, which is similar to the nematic director field \nofr~but with the identification
\tetr~$\equiv$~ -\tetr~$\equiv$~$\mathbf{\hat{e}}_z$ $\times$~\tetr~$\equiv$ - $\mathbf{\hat{e}}_z$ $\times$ \tetr~ of both parallel and perpendicular orientations ($\mathbf{\hat{e}}_z$ is the unit vector perpendicular to the two-dimensional system).
The topological charge $\Qtet$ of a tetratic defect is always a multiple of $1/4$ and can be defined analogously to Eq.~\eqref{eq_Q} as
\begin{align}
\label{eq_q}
\Qtet = \frac{1}{2\pi} \oint_{\mathcal{C}(\kappa)}~\left [\mathbf{\hat{t}} (\kappa) \times \frac{\partial \mathbf{\hat{t}} (\kappa)}{\partial \kappa} \right ] d\kappa\,,
\end{align}
where the closed contour $\mathcal{C}$ may traverse grain boundaries, which is forbidden for a nematic director field.
{\rene This allows us to} identify two isolated pointlike tetratic defects at the end points of a grain boundary.
{\rene For example, the contours $\mathcal{C}_2$ and $\mathcal{C}_3$ in Fig.~\ref{fig_new_contours} enclose a tetratic bulk defect and a virtual defect at the boundary, respectively.}

As a third step, we apply the same continuum assumption to networks of grain boundaries, which reveals the location of additional tetratic point defects at the nodes, which represent a junction point of three grain boundaries, {\rene compare, e.g., the contour $\mathcal{C}_4$ in Fig.~\ref{fig_new_contours}.}
 For the convex confinement shapes considered in our work,
 we typically observe defect networks of total nematic charge $\Qnem=1/2$, 
 consisting of $n$ nodes and $n+2$ end points with tetratic charge $\Qtet=-1/4$ and $\Qtet=+1/4$, respectively.
 This decomposition reflects the fact that the nematic charge
 \begin{align}
  \Qnem=  \sum_j \Qtet_j 
    \label{eq_conservationTET}
\end{align}
of a defect equals the sum of all tetratic point charges $\Qtet_j$ enclosed by the same contour, {\rene compare, e.g., the contour $\mathcal{C}_5$ in Fig.~\ref{fig_new_contours}.}
Formally, the topological charge of a grain-boundary network is thus exclusively located at its nodes and end points.
Globally, the fundamental conservation laws, Eq.~\eqref{eq_conservationlaw}, for the total nematic charge $\Qnem$ and accordingly $\sum_i\Qtet_i=1$ 
for the total tetratic charge $\Qtet$ can be
nicely illustrated in terms of the two types $A$ (with $\Qnem=1/4+1/4=1/2$) and $B$ (with $\Qnem=1/4-1/4=0$) of grain-boundary lines, introduced in the main text
as building blocks of the emerging defect structures:
we usually identify two $A+nB$ networks, which consist of one type-$A$ and $n$ type-$B$ defect, and a number of standalone type-$B$ grain boundaries.

{\rene As a final step, we must decide for which of the observed grain boundaries the continuum assumption is justified
and how to classify the defect structures in the case it is not.
This issue is dealt with in the following section.}

\begin{figure}[t]
\begin{center}
\includegraphics[width=1.00\linewidth]{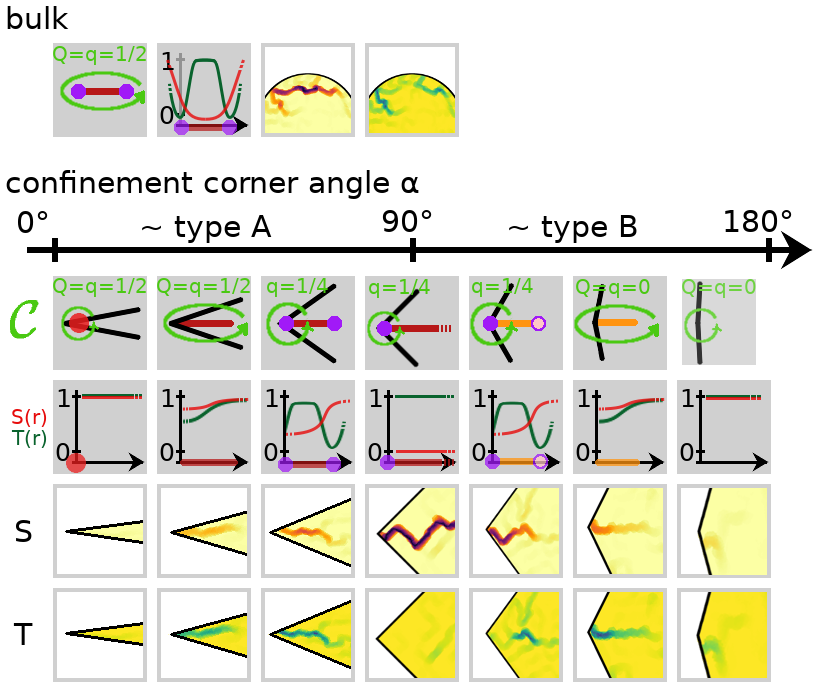}

\caption{\rene
Schematic illustration of the behavior of the nematic order parameter $S$ (red curves) and the tetratic order parameter $T$ (green curves)  along a grain-boundary line (horizontal coordinate in the plots) together with selected fields extracted from the simulation data representing the indicated scenario.
We also draw in each case a contour $\mathcal{C}$ (green arrow) that encloses a representative nematic charge $\Qnem$ and/or tetratic charge $\Qtet$.
\textbf{Top:} type-$A$ bulk defect, typically found in circular confinement.
\textbf{Bottom:} the typical defect type emerging at a corner of the confining wall depends on the opening angle $\alpha$. If no tetratic point charges are drawn at the endpoints of a grain boundary, we consider these as partially annihilated. Further details are given in the text.
}
\label{fig_angles}
\end{center}
\end{figure}

\section{Detection of tetratic defects in frustrated smectics}

\subsection{Local fields of the orientational order parameters \label{sec_fields}}

 Having introduced a continuum classification of smectic grain boundaries in Sec.~\ref{sec_topSM}, we now address the defect structure arising from frustrated smectic order in extreme confinement as represented by our simulation data.
One way to visualize the emerging topological defects is
through the local profiles of the nematic order parameter $S$ and the tetratic order parameter $T$.
If a grain-boundary line is located in the bulk, i.e., not attached to the boundary,
its topological structure is unambiguously revealed by these fields.
{\rene As emphasized in Fig.~\ref{fig_angles},}
the lines along which the $S$ field is close to zero are clearly visible, %
whereas the tetratic signal in the middle of the line is not distinguishable from the background.
However, the $T$ field is low at both ends of the line,
indicating the two end-point defects. %
If a grain-boundary network is connected to the wall, 
both the strength of the signal in the order parameters (deviation from the bulk value one) along a grain boundary
and the observed type of the free-standing defects typically depends on the opening angle $\alpha$ of the adjacent corner,
{\rene as illustrated in Fig.~\ref{fig_angles} and detailed in the following}.
Corners with $\alpha\simeq 90^\circ$ induce ideal grain boundaries with with $S\simeq0$
and promote perfect tetratic ordering throughout the system.
The tetratic end point defects are fully located on the system boundary and can therefore not be resolved in the $T$ field.
{\rene Departing from the particular case $\alpha\simeq 90^\circ$, the tetratic order along a grain boundary is also frustrated close to the confining wall.
For $45^\circ\lesssim\alpha\lesssim90^\circ$ and $90^\circ\lesssim\alpha\lesssim135^\circ$ the nematic signal along a grain-boundary line opposes the tetratic signal ($S$ increases when $T$ decreases and vice versa).
For both $0^\circ\lesssim\alpha\lesssim45^\circ$ and $135^\circ\lesssim\alpha\lesssim180^\circ$ 
the tetratic charges no longer provide isolated signals, i.e., they are partially merged to a line defect
whose order parameter field qualitatively resembles the nematic one.
Such a grain boundary does not accommodate perpendicular smectic layers,  compare, e.g., the contour $\mathcal{C}_6$ in Fig.~\ref{fig_new_contours}}.
{\rene In the limit $\alpha\rightarrow180^\circ$, adjacent smectic layers are parallel 
and there is no topological defect.
This corresponds to a type-$B$ grain boundary, typically observed for $\alpha \gtrsim 90^\circ$, whose oppositely charged tetratic endpoints are completely annihilated.
In the opposite case, $\alpha\rightarrow0^\circ$, there is a point defect of charge $\Qnem=\Qtet=1/2$ in the corner, corresponding to the 
two merged endpoints of a type-$A$ defect, generally preferred for $\alpha\lesssim 90^\circ$.}

{\rene 
The given thresholds for the opening angle $\alpha$ reflect the different symmetry inherent to the nematic and tetratic order.
The resulting intuition helps to predict
the expected type of defect induced by a corner.
In particular if there is only a single grain-boundary line, 
the information in the order-parameter field can be used as a basis for the decision
whether or not to assign isolated tetratic end points to the defect, as indicated in Fig.~\ref{fig_angles}.
Regarding, however, the variety of possible defect structures laid out in Sec.~\ref{sec_rel_freq},
we take a cue from our domain criterion described in Sec.~\ref{sec_analysis}
and attach a branch of type $B$ to a grain-boundary network only if it separates two different domains.
Otherwise, we consider the endpoints of such a building block as annihilated and no such branch is drawn into our schematic defect structures.}

\subsection{Tetratic defect analysis based on smectic layers}

\begin{figure}[t]
\begin{center}
\includegraphics[width=0.9800\linewidth]{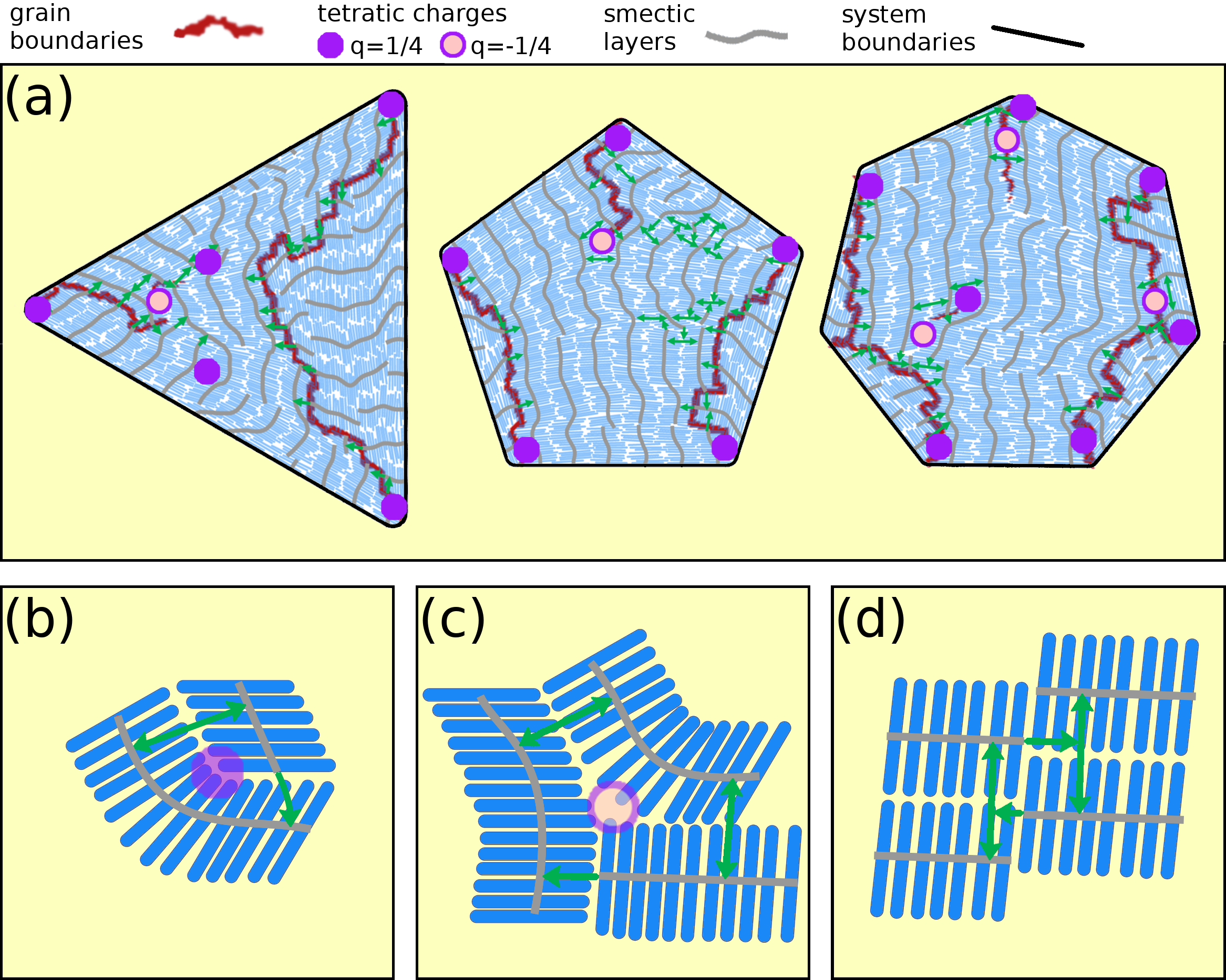}

\caption{\rene Detecting tetratic charges based on smectic {\rene layers determined by the algorithm outlined in Sec.~\ref{sec_analysis}.} 
\textbf{a)} Exemplary distribution of tetratic point charges
in three simulation snapshots. 
\textbf{b-d)} Illustration of how to identify different tetratic charges. The gray lines indicate the detected smectic layers. Green arrows with one or two tips are added manually to indicate connections between these layers in the vicinity of the topological defects. Then the tetratic charge $q=1-t/4$ can be read of from the number $t$ of tips. Further details are given in the text.}
\label{fig_algorithmic_defects}
\end{center}
\end{figure}

 To provide an alternative point of view on tetratic defects in frustrated smectics, we focus our analysis on the smectic layers, instead of the particle-resolved picture in the previous section.
 Our main goals are to provide a deterministic distinction between isolated and (partially) annihilated defects and to explicitly detect boundary defects.
In the frustrated smectic configurations we consider here, many of the smectic layers are tilted to some degree, and as a result the particles are not aligned perpendicular to the direction of the smectic layer. Hence, the difference in nematic orientation across a grain boundary can be significantly different from $\pi / 2$, resulting in a low local tetratic order parameter near the grain boundary. As a result, pinpointing defects based on the tetratic order parameter is not easy in practice.
We thus require a more robust way of identifying the topological defects
associated with the tetratic order in frustrated smectic liquid crystals from our simulation data.

To circumvent the difficulties associated with the local order parameter field, we determine the 
directions and connectivity of the smectic layers. Specifically, we consider two particles to be in the same smectic layer based on a bonding criterion that takes into account the positions and orientations of pairs of particles, see Sec.\ref{sec_analysis} for details. We then draw a smoothened center line along the length of each layer to mark its direction. The smoothening is performed by first determining an overall direction for the smectic layer (taken to be perpendicular to the nematic director of all particles in the layer), and then applying a low-pass filter to the component of the particle positions perpendicular to this director.  
We can now use these lines to guide us in identifying the behavior of the tetratic director as we move through the system. Along the center line of the smectic layers, the orientation of the tetratic director varies smoothly. 
Moreover, we can connect two lines by green arrows that are either parallel, or meet at an approximately perpendicular junction, such that the tetratic director can be assumed to vary smoothly along this connection.
The resulting closed lines allow us to locate the defects in the tetratic order, as illustrated in Fig.~\ref{fig_algorithmic_defects}. 
To do this, we consider two types connections between smectic layers. 
First, between two parallel smectic layers within the same smectic domain, we can draw direct connections perpendicular to the two smectic layers, denoted by double-headed green arrows. 
 Second, when a smectic layer terminates against the side of another layer, such that the last particle in the first layer lies parallel to the smectic direction of the second layer, we draw a single-headed arrow,
 which by definition crosses an area of low nematic order.

As shown in Fig.~\ref{fig_algorithmic_defects}, it is then possible to identify a topological charge $q=1-t/4$ of tetratic defects ,
by counting the number $t$ of turns represented by the tips of the green arrows. In an idealized interpretation of any loop, switching direction at the tip of a green arrow is a turn of $90^\circ$, which has no effect on the tetratic order. Following the closed loop in Fig.~\ref{fig_algorithmic_defects}b, we encounter three sharp corners of $90^\circ$ (with no effect on the tetratic order) plus the smooth rotation of the smectic director by a final $90^\circ$, completing the loop. This quarter-turn of the tetratic director indicates the presence of a defect with $q=+1/4$ topological charge. Similarly, closed loops with 5 turns at a right angle, like in Fig.~\ref{fig_algorithmic_defects}c must carry a topological charge of $q=-1/4$.
Finally, we should consider non-topological defects that are common in our configurations. These can be either dislocations or simply sudden shifts in a smectic layer perpendicular to its smectic direction, and do not significantly impact the orientations of the particles: they occur in regions of high nematic order. An example is shown in Fig.~\ref{fig_algorithmic_defects}d, with connections between the different layers drawn to show that no topological defects occur, as the closed loop has exactly four sharp corners of $90^\circ$.

We can now apply this approach to the simulated configurations, as illustrated in Fig.~\ref{fig_algorithmic_defects}a. For this, it is convenient to include the nematic order parameter field as a background. In particular, as terminations of one smectic layer against another (single-headed green arrows) only occur in regions of low nematic order, in practice we only have to focus on the regions close to the grain boundaries to find tetratic defects.  In each figure, we draw only the connections required to identify the topological defects in the system.  Then, additional connections between parallel layers are added to narrow down the regions where defects are located. 
Moreover, as detailed in Sec.~\ref{sec_topSM}, we assume the tetratic director field to vary smoothly along the outside wall. To reflect this, we draw a smoothened outline of the system boundary in each image. Smectic layers can terminate at the wall, providing an additional connection that should again be interpreted as a rotation by $90^\circ$ in any closed loop that includes it. 
In other words, the wall represents a double-headed arrow (which is not drawn for presentation reasons).

We stress that identifying tetratic defects in this way is relatively straightforward. Arrows are drawn manually, but there is usually no freedom of choice: when a smectic layer terminates in a region of low nematic order, it is connected to the perpendicular layer just ahead of it. If no such layer is immediately visible, then occasionally another connection in this region should be made first to allow the first layer to terminate by drawing an arrow which ends on another perpendicular arrow (see, e.g., on the right-hand side of the heptagon in Fig.~\ref{fig_algorithmic_defects}a.
The only freedom of choice in this method occurs when it is not clear whether two adjacent smectic lines should be considered perpendicular or parallel. For example, one could imagine continuously deforming the region around the defect pair just to the left of the center of the heptagon in Fig.~\ref{fig_algorithmic_defects}a to be more similar to an edge dislocation, as in Fig.~\ref{fig_algorithmic_defects}d, which would result in the annihilation of the two defects.
However, note that any choice leaves the overall topological charge of the system intact, as one would expect.

 Compared to the analysis of order-parameter fields, described in Sec.~\ref{sec_fields}, our layer-based approach generally resolves a larger number of defect pairs, albeit
these are usually close to annihilation. For example, 
the distribution of tetratic charges in the free-standing grain boundary 
at the top corner of the heptagon in Fig.~\ref{fig_algorithmic_defects}a
corroborates our interpretation of partially annihilated endpoint charges.
In conclusion, while both methods provide a solid understanding of the tetratic topology in their own right, the algorithm presented above is perhaps most suitable
to track the defect motion in a dynamical setup.}

\section{Relative frequencies of observed liquid crystal structures}
\label{sec_rel_freq}

\begin{figure*}[t]
\begin{center}
\includegraphics[width=0.8125\linewidth]{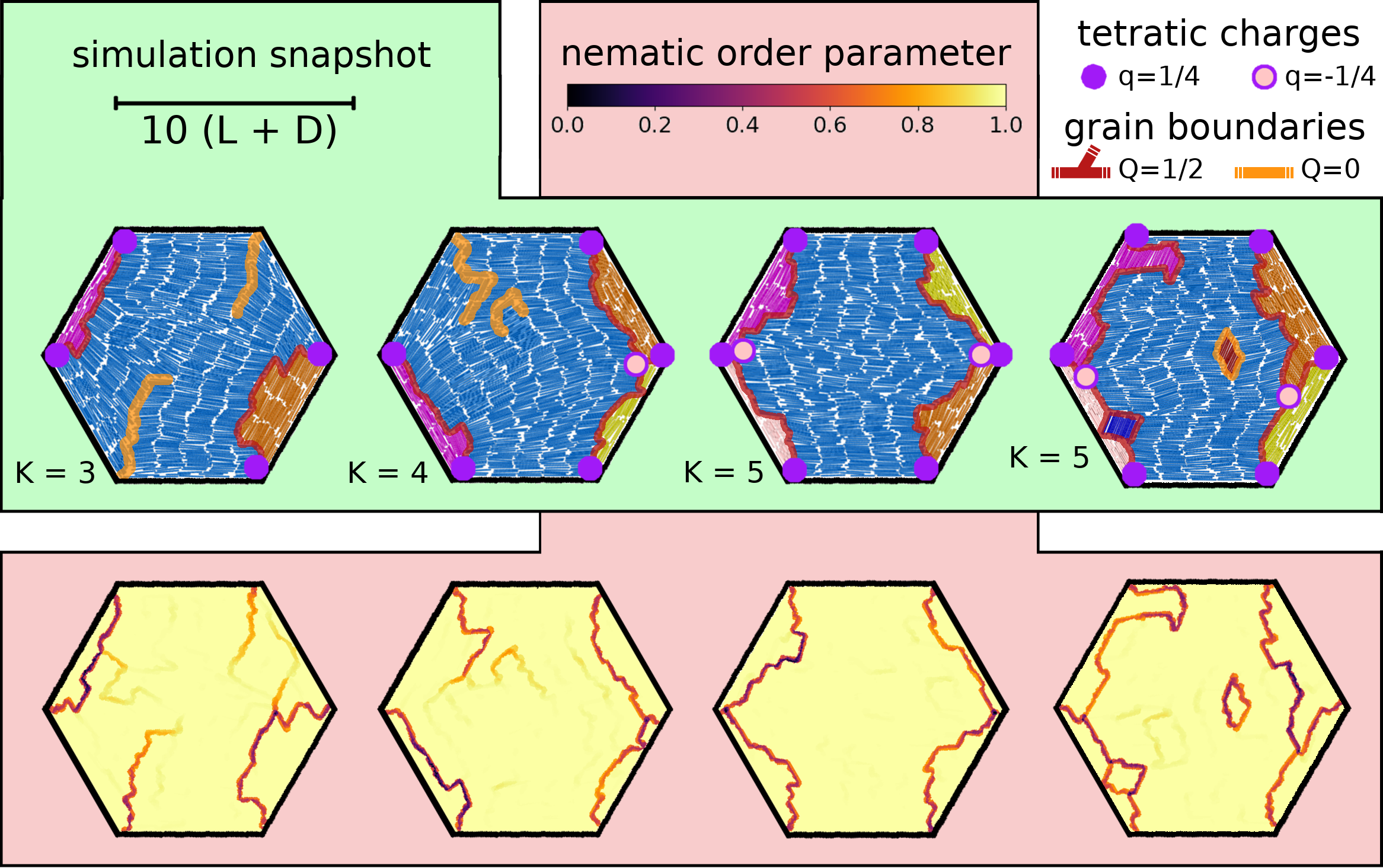}
\caption{Comparison of representative bridge states with $K$ domains as labeled from simulations of $N=1000$ hard rods confined to a hexagonal cavity. The corresponding relative frequencies are given in Tab.~\ref{tab_rel_freq_table}. 
The structure on the right illustrates that we do not count singular domains located in the middle of the cavity, as described in the caption of Tab.~\ref{tab_rel_freq_table}.
\textbf{Top row:} particle snapshots with superimposed grain-boundary networks (compare Fig.~2 in main manuscript).
\textbf{Second row:}
nematic order parameter field \Sofr.}

\label{fig_hex_dom_range}
\end{center}
\end{figure*}

In general, each smectic structure in simply-connected confinement features a total topological charge $\Qnem=\Qtet=1$ typically 
distributed over two $A+nB$ grain-boundary networks (containing one type-$A$ and $n$ type-$B$ building blocks each) 
and a number of additional type-$B$ grain boundaries. 
To identify the most likely network composition, we differentiate between the occurring defect structures by counting the number of domains, according to the analysis in Sec.~\ref{sec_analysis}. Considering hexagonal confinement as an exemplary case,
\figref{fig_hex_dom_range} shows 
how the number of domains relates to the complexity of the emerging networks: when completely connected to the wall, an $A + nB$ network separates $(n+2)$ domains. 

In Tab.~\ref{tab_rel_freq_table}, we list the relative frequencies for structures labeled according to the number of domains for a range of polygonal confinements. 
For polygons with smaller numbers of corners, we observe that the average number of domains $\overline{\mathbf{K}}$ increases with increasing number of corners, peaking for 
dodecagon (12 corners) and tridecagon (13 corners).
Further increasing the number of corners, $\overline{\mathbf{K}}$ decreases again, as the confinement gradually approaches a circular shape. This behavior reflects a general trend, which is also reflected by the most frequently observed defect structure for a given confinement. 
The particular distribution for a given polygon depends in detail on the ratio of the side length (determined by the fixed particle number and area fraction) to the length of the rods. For instance, the triacontagonal (30 corners) confinement, for which this ratio is equal to $1.07$, displays a relatively large number (compared to 40 corners) of single-domain structures, which can be related to the high probability of a continuous string of rods around the perimeter.

\begin{table}[t]
\begin{center}
\begin{tabular}[t]{l|c|c|c|c|c|c|c|c|c|c|c|r}
 \textbf{confinement}   & \mbox{} & \multicolumn{10}{c}{\textbf{number of domains $\mathbf{(K)}$}}&\\
\textbf{(corners)} & $\overline{\mathbf{K}}$ & $\mathbf{1}$ & $\mathbf{2}$ & $\mathbf{3}$ & $\mathbf{4}$ & $\mathbf{5}$ & $\mathbf{6}$ & $\mathbf{7}$ & $\mathbf{8}$ & $\mathbf{9}$ & $\mathbf{10}$ & $\mathbf{11}$ \\
\hline

circle ($0$, $\infty$) & $1.64$ &  $\mathbf{0.55}$ & $0.10$ & $0.03$ & $0.01$ & \mbox{} & \mbox{} & \mbox{} & \mbox{} & \mbox{} & \mbox{}& \mbox{}\\




triangle (3) & $2.34$ & $0.08$ & $\mathbf{0.53}$ & $0.35$ & $0.03$ & \mbox{} & \mbox{} & \mbox{} & \mbox{} & \mbox{} & \mbox{}  & \mbox{}\\



square (4)& $2.82$ & \mbox{} & $0.21$ & $\mathbf{0.75}$ & $0.03$ & \mbox{} & \mbox{} & \mbox{} & \mbox{} & \mbox{} & \mbox{} & \mbox{} \\


pentagon (5) & $3.52$ & \mbox{} & $0.03$ & $0.45$ & $\mathbf{0.49}$ & $0.03$ & \mbox{} & \mbox{} & \mbox{} & \mbox{} & \mbox{}  & \mbox{}\\



hexagon (6)& $4.72$ & \mbox{} & \mbox{} & $0.05$ & $0.21$ & $\mathbf{0.68}$ & $0.05$ & \mbox{} & \mbox{} & \mbox{} & \mbox{}& \mbox{} \\


heptagon (7)& $5.16$ & \mbox{} & \mbox{} & $0.01$ & $0.16$ & $\mathbf{0.54}$ & $0.25$ & $0.04$ & \mbox{} & \mbox{} & \mbox{}& \mbox{} \\


octagon (8) & $5.49$ & \mbox{} & \mbox{} & \mbox{} & $0.05$ & $\mathbf{0.53}$ & $0.32$ & $0.09$ & $0.02$ & \mbox{} & \mbox{}& \mbox{} \\


nonagon (9) & $6.17$ & \mbox{} & \mbox{} & \mbox{} & $0.02$ & $0.20$ & $\mathbf{0.46}$ & $0.26$ & $0.06$ & $0.01$ & \mbox{} & \mbox{} \\



decagon (10) & $ 6.25$ & \mbox{} & \mbox{} & \mbox{} & $0.02$ & $0.18$ & $\mathbf{0.40}$ & $0.33$ & $0.06$ & $0.01$ & \mbox{}& \mbox{} \\



hendecagon (11) & $6.76$ & \mbox{} & \mbox{} & \mbox{} & $0.02$ & $0.10$ & $0.27$ & $\mathbf{0.36}$ & $0.20$ & $0.05$ & \mbox{} & \mbox{} \\



dodecagon (12)& $7.04$ & \mbox{} & \mbox{} & \mbox{} & $0.02$ & $0.08$ & $0.22$ & $\mathbf{0.35}$ & $0.22$ & $0.08$ & $0.03$ & $0.01$ \\



tridecagon (13) & 7.05& \mbox{} & \mbox{} & $0.01$ & $0.03$ & $0.09$ & $0.21$ & $\mathbf{0.30}$ & $0.25$ & $0.10$ & $0.02$ & \mbox{} \\



tetradecagon (14) & $6.46$ & $\mbox{}$ & $\mbox{}$ & $0.02$ & $0.05$ & $0.16$ & $0.27$ & $\mathbf{0.28}$ & $0.26$ & $0.05$ & $0.01$ &  $\mbox{}$\\



pentadecagon (15)& $5.71$ & $\mbox{}$ & $0.02$ & $0.05$ & $0.13$ & $0.23$ & $\mathbf{0.26}$ & $0.20$ & $0.08$ & $0.03$ & $\mbox{}$ & $\mbox{}$\\



hexdecagon (16) & $2.97$ & $0.14$ & $\mathbf{0.28}$ & $0.26$ & $0.16$ & $0.10$ & $0.04$ & $0.01$ & \mbox{} & \mbox{} & \mbox{} & \mbox{} \\



icosagon (20) & $2.43$ & $0.23$ & $\mathbf{0.36}$ & $0.24$ & $0.11$ & $0.05$ & $0.01$ & \mbox{} & \mbox{} & \mbox{} & \mbox{} & \mbox{} \\



triacontagon (30) & $1.09$ & $\mathbf{0.91}$ & $0.08$ & \mbox{} & \mbox{} & \mbox{} & \mbox{} & \mbox{} & \mbox{} & \mbox{} & \mbox{} & \mbox{} \\



tetracontagon (40) & $1.88$ & $\mathbf{0.44}$ & $0.34$ & $0.16$ & $0.05$ & $0.02$ & \mbox{} & \mbox{} & \mbox{} & \mbox{} & \mbox{} & \mbox{} \\

\end{tabular}
    \caption{Average number of domains $\overline{\mathbf{K}}$ and relative frequencies of structures
    with $\mathbf{K}$ domains occurring in our simulations, rounded to two decimal places.  The most frequent structure is marked by a boldfaced number. The data are sampled from 1000 simulations for each confinement. For this statistic, we only count domains which align with at least one wall to better reflect the typical defect geometry of the confinement, compare \figref{fig_hex_dom_range}.}
\label{tab_rel_freq_table}
\end{center}
\end{table}

\section{Additional simulation data}
 
The distinctive topology of the defects, we observe in our confined systems, is influenced by various quantities, such as area fraction $\eta_2$, geometry of the particles (aspect ratio $p$), as well as confining geometry and system size (particle number $N$). To further explore these different aspects, we present in the following sets of simulation data generated with the protocol described in Sec.~\ref{sec_simproc}.
First, we consider additional geometries for the simulation parameters $N=1000$, $p=15$ and $\eta_2=0.75$ used in the main text.
In detail, we exemplify the dependence of the typical length of a grain boundary on the sharpness of the adjacent corner by considering a range of isosceles triangles in Sec.~\ref{sec_isotriangles}
and discuss the effects of a varying local curvature of the confining walls by the example of different rounded triangles in Sec.~\ref{sec_rtria}.
We then elaborate on the dependence of the topological defect structure on the system size in Sec.~\ref{sec_size}. Finally, to shed light on the relation between simulation and experiment, we explore a range of area fractions and aspect ratios in Sec.~\ref{sec_exp}.

\subsection{Isosceles triangles \label{sec_isotriangles}} 

\begin{figure*}[!!!!!!!!!!!!!!!!!!!p]
\begin{center}
\includegraphics[width=\linewidth]{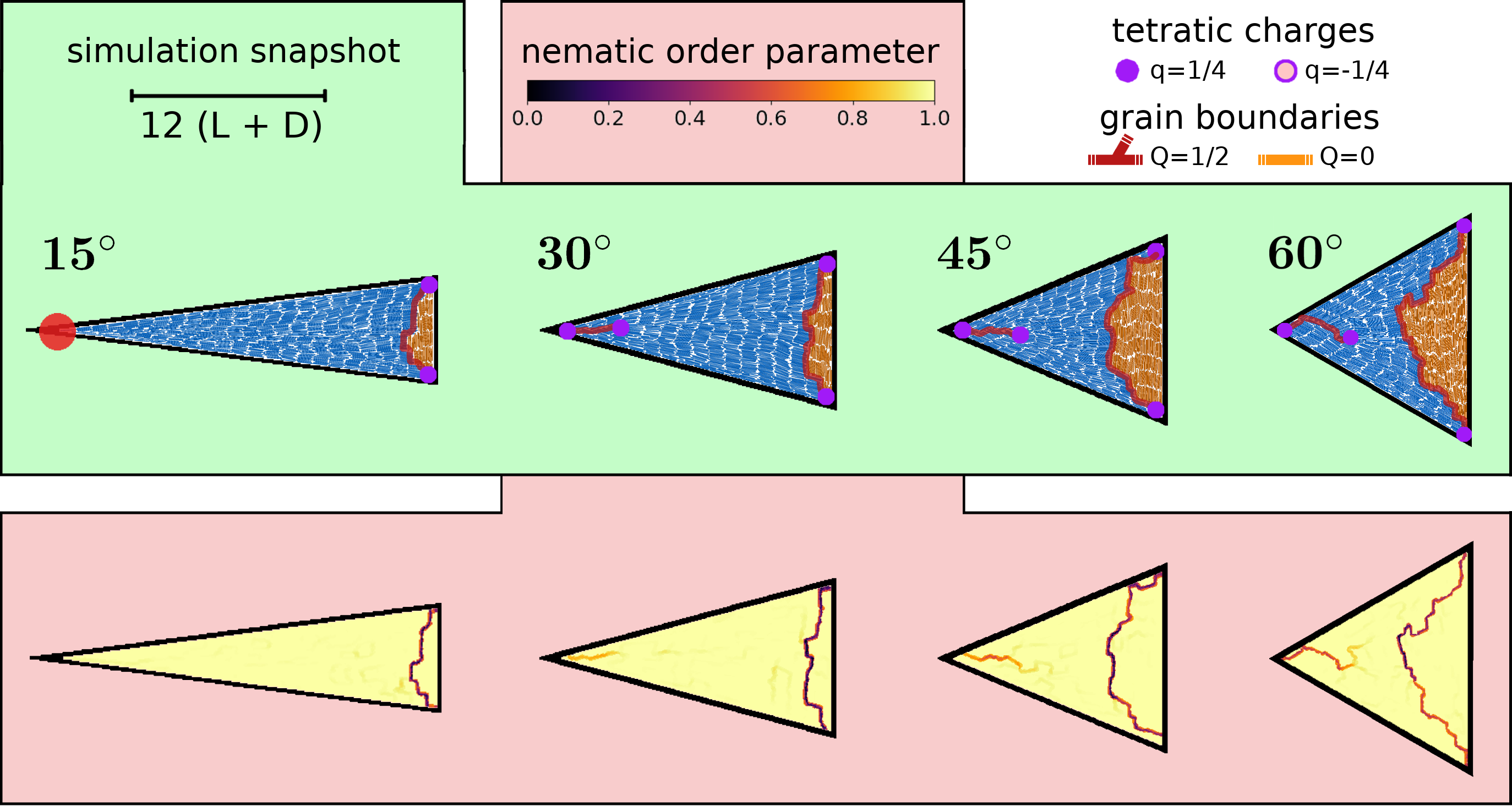}
\caption{Topological defect structure of representative simulation results for hard rods  with aspect ratio $p=15$ confined to isosceles triangular cavities with tip angles (as labeled), 
where the structure shown at $60^{\circ}$ coincides with that 
in Fig.~3 in the main manuscript. Particle snapshots and orientational order parameter field \Sofr~as in \figref{fig_hex_dom_range}.}
\label{fig_isosc_triangles}
\end{center}\vspace*{0.2cm}
\begin{center}
\includegraphics[width=\linewidth]{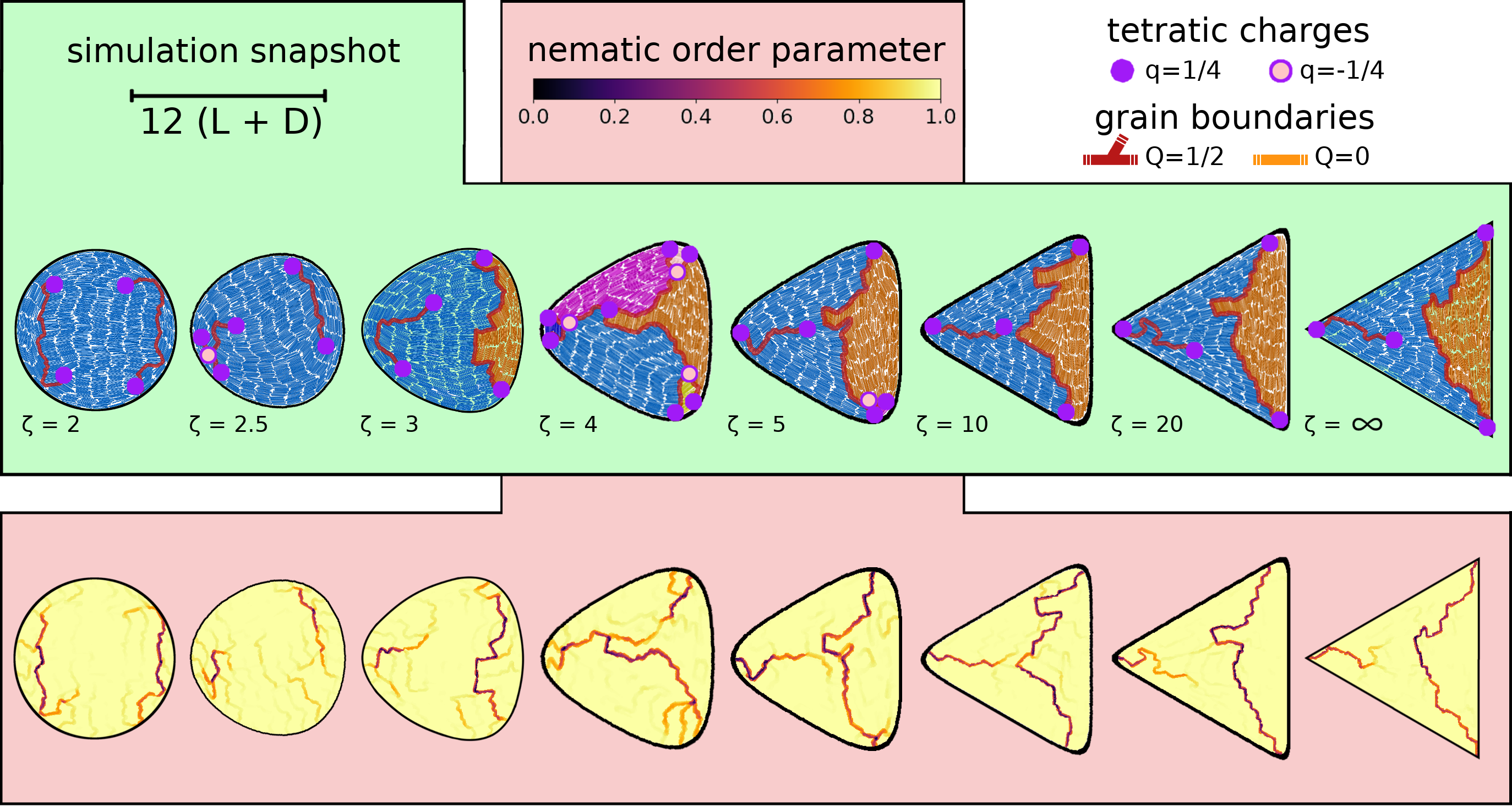}
\caption{Topological defect structures of representative simulation results for $N=1000$ hard rods  with aspect ratio $p=15$ confined to triangular cavities with varying roundness $\zeta$, defined in Eq.~\eqref{eq_rounded_tria_eq}, as labeled. The structures with $\zeta\in\{2,~3,~\infty\}$ coincide with those presented in Fig.~3 in the main manuscript.
\smash{Particle snapshots and orientational order parameter field \Sofr~as in \figref{fig_hex_dom_range}.}}
\label{fig_rounded_tria_raster}
\end{center}
\end{figure*}

To study the effect of the of the sharpness of a corner on the typical length of the adjacent grain boundary, we consider in \figref{fig_isosc_triangles} a range of sharp isosceles triangles.
The appearance of the free-standing type-$A$ defect, typically attached to the sharpest corner of the polygons, critically depends on the opening angle, as {\rene illustrated in \figref{fig_angles} and} depicted in \figref{fig_isosc_triangles}. 
As elaborated in the main manuscript, the free-standing grain boundary in equilateral triangles tends to protrude deeply into the bulk. For intermediate opening angles, we typically observe that the length of the defect decreases compared to the legs of the triangle and the degree of orientational frustration becomes lower {\rene as the tetratic endpoints begin to merge}. For angles $\lesssim15^{\circ}$ we observe an isolated perfect point defect with $\Qnem=\Qtet=+1/2$ as smectic layering is suppressed by the cramped geometry in the vicinity of the tip.

\begin{figure*}[b]
\begin{center}
\includegraphics[width=1.0\linewidth]{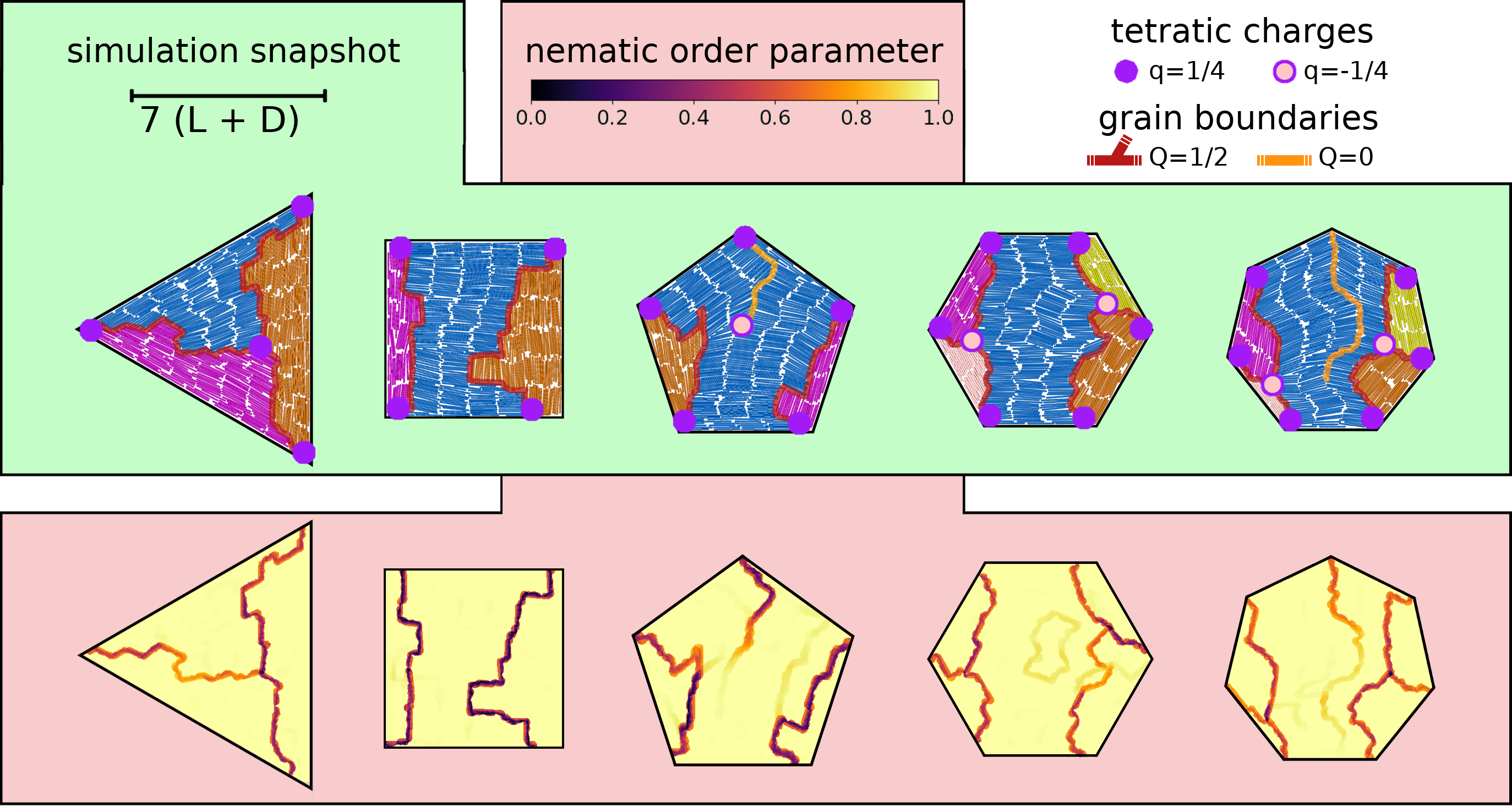}
\caption{Topological defect structure of representative simulations for $N=500$ hard rods with aspect ratio $p=15$ in a range of regular polygons. 
Particle snapshots and orientational order parameter field \Sofr~as in \figref{fig_hex_dom_range}.}
\label{fig_smaller_systems}
\end{center}
\end{figure*}

\subsection{Rounded triangles}
\label{sec_rtria}

As mentioned in the main manuscript, the curvature of the outer system boundary plays a central role in the formation and resulting topological interpretation of grain-boundary networks. 
To elaborate on the dependence of the topological defect structure on the curvature 
of confining cavities, we consider rounded polygons generally defined by the equation
\begin{align}
    \label{eq_rounded_tria_eq}
    1 = \left [ \sum^M_k  \left (  \mathbf{r} \cdot  \mathbf{b}_k   - \mathcal{W}  \right )^\zeta    \right ]^{1/\zeta}
\end{align}
with $\mathbf{r} \in \mathbb{R}^2$, $\mathbf{b}_k = \left ( \cos(k2\pi/M) , \sin(k2\pi/M) \right )$ and $\mathcal{W} = \tan(\pi/(2M))\sin(2\pi/M)$,
such that $M$ and $\zeta$ denote the number and sharpness of the corners, respectively. In particular, $\zeta = 2$ defines a circle and the limit $\zeta \rightarrow \infty$ corresponds to sharp corners.

\begin{figure*}[!!!!!!!!!!p]
\begin{center}
\includegraphics[width=1.0\linewidth]{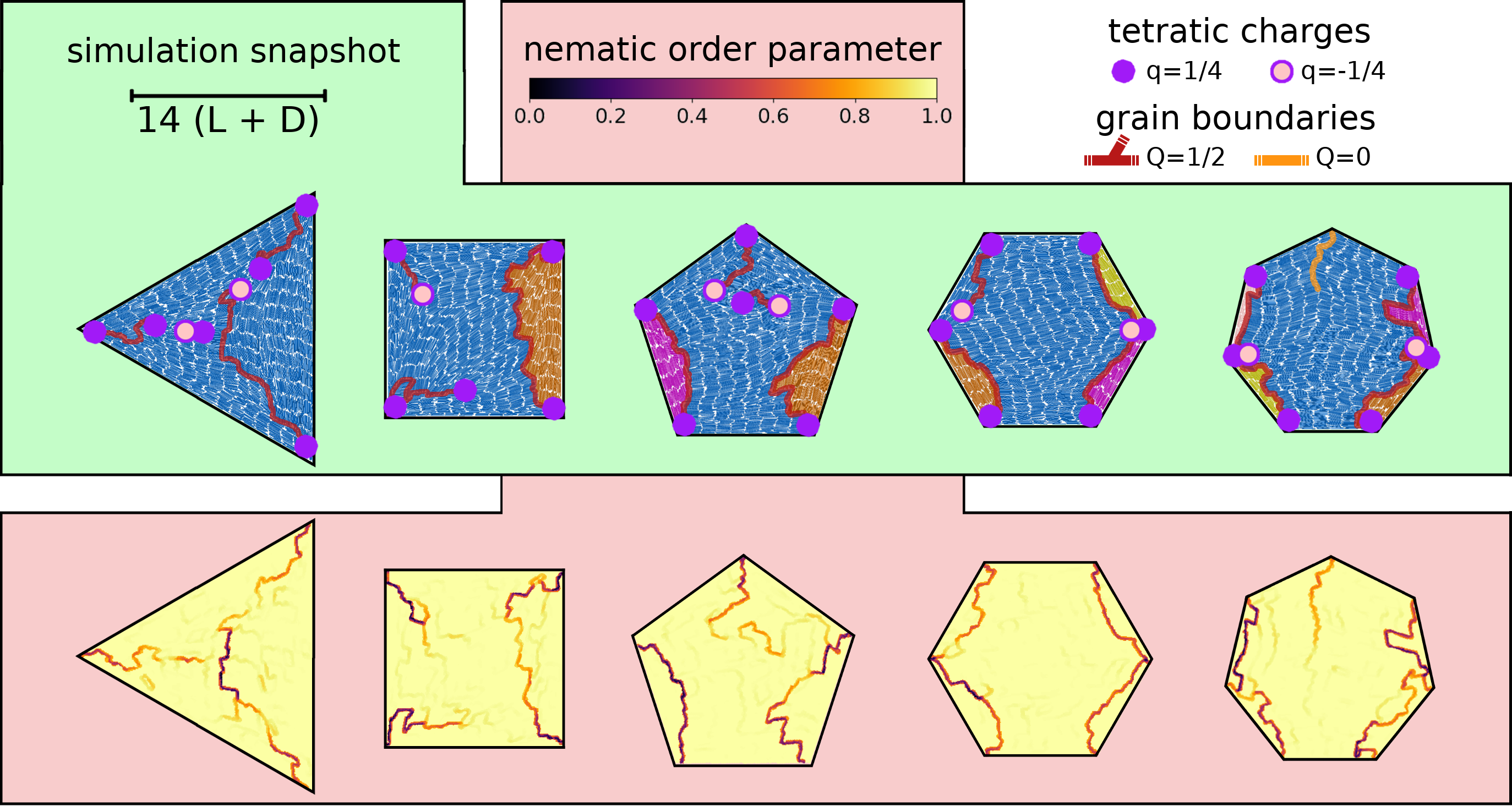}
\caption{Topological defect structure of representative simulations for $N=2000$ hard rods with aspect ratio $p=15$ in a range of regular polygons. 
Particle snapshots and
orientational order parameter field \Sofr~as in \figref{fig_hex_dom_range}.}
\label{fig_imd_systems}
\end{center}\vspace*{0.2cm}

\begin{center}
\includegraphics[width=1.0\linewidth]{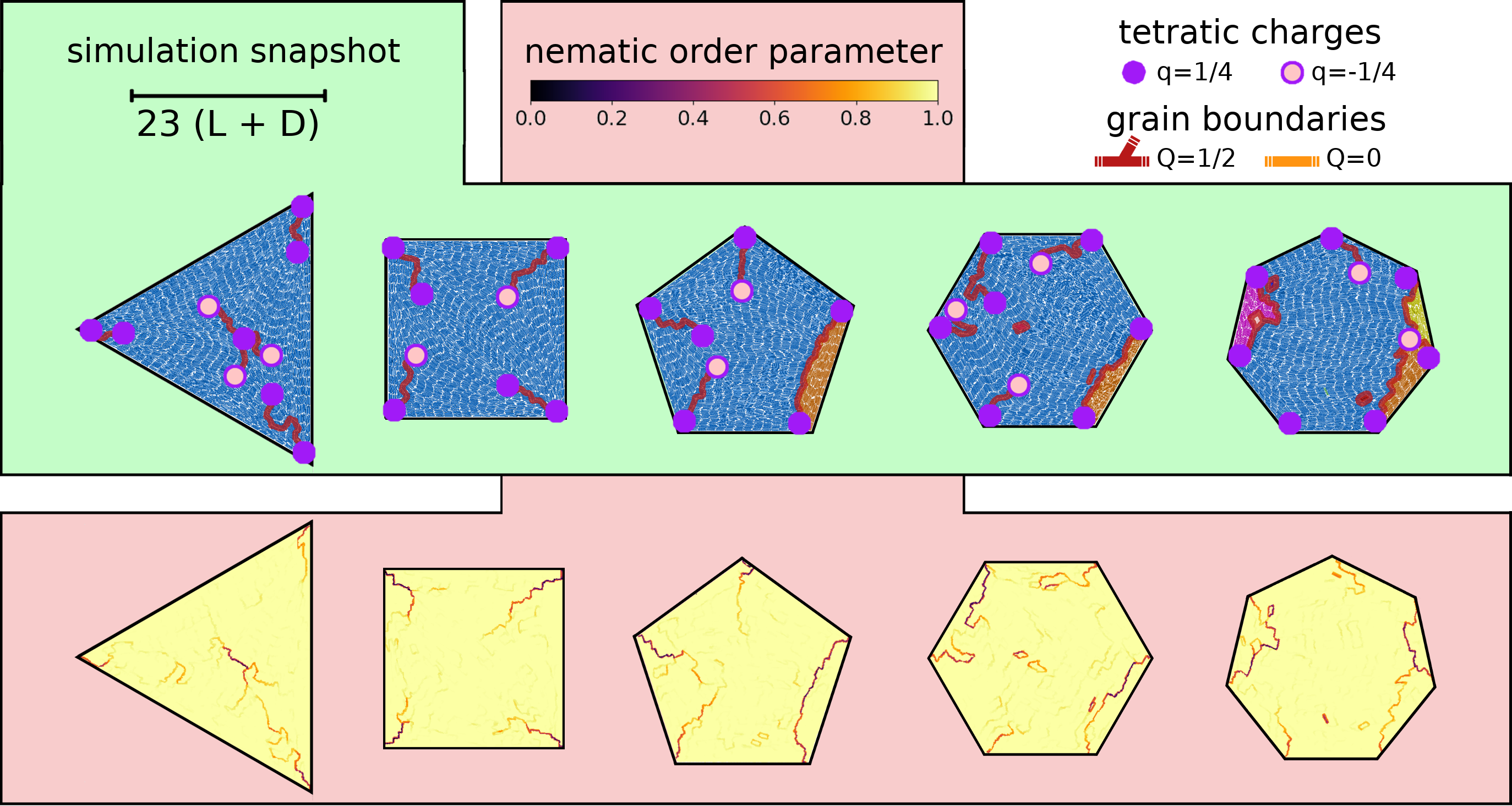}
\caption{Topological defect structure of representative simulations for $N=5000$ hard rods with aspect ratio $p=15$ in a range of regular polygons. 
Particle snapshots and
orientational order parameter field \Sofr~as in \figref{fig_hex_dom_range}.}
\label{fig_bigger_systems}
\end{center}
\end{figure*}

Focusing on $M=3$, \figref{fig_rounded_tria_raster} compares the typical structure for a large range of rounded triangles. 
For sharp corners with infinite local curvature, the sudden shift in the orientation of the wall causes a frustration of the orientational order directly at the walls, resulting in grain boundaries attached to each corner.
The resulting defect structure with two clearly separated domains remains dominant for rounded triangles with $\zeta\gtrsim4$ throughout.
For $\zeta\lesssim4$, the dominant structure resembles that in circular confinement, characterized by a single domain and two opposite grain-boundary lines on each side of the confinement, which tend to align locally with the confining wall. The degree of attachment to the system boundary, generally decreases with $\zeta$, {\rene which demonstrates the convenience of our virtual treatment of boundary defects}.

For intermediate values of the roundness parameter around $\zeta \approx4.0$, where the curvature 
radius at the corners is comparable to the rod length, we find additional structural elements: the grain boundaries tend to branch out close to the corners, resulting in small domains incorporating one to two smectic layers. Moreover, in this parameter range, we find structures that feature either one, two or three large domains, reflecting a strongly fluctuating connectivity of the defect networks.


\begin{table}[t]
\begin{center}
\begin{tabular}[t]{l|c|c|c|c|c|c|c|c|r}
 \mbox{} & \mbox{} & \mbox{} & \multicolumn{6}{c}{\textbf{number of domains $\mathbf{(K)}$}}&\\
 $N$ & sim-\# &  $\overline{\mathbf{K}}$ & $\mathbf{1}$ & $\mathbf{2}$ & $\mathbf{3}$ & $\mathbf{4}$ & $\mathbf{5}$ & $\mathbf{6}$ & $\mathbf{7}$ \\
\hline




$500$ & $1000$ & $4.85$ & \mbox{} & \mbox{} & $0.03$ & $0.23$ & $\mathbf{0.62}$ & $0.11$ &  $0.01$ \\


$1000$ &  $1000$ & $4.72$ & \mbox{} & \mbox{} & $0.05$ & $0.21$ & $\mathbf{0.68}$ & $0.05$ & \mbox{} \\



$1500$ & \rene $1000$ & $4.62$ & \mbox{} & $0.01$ & $0.06$ & $0.33$ & $\mathbf{0.50}$ & $0.09$ & $0.01$ \\



\rene $1777$ & \rene $300$ & $3.74$ & $0.01$ & $0.12$ & $0.25$ & $\mathbf{0.38}$ & $0.22$ & $0.02$ &  \mbox{} \\

$2000$ & \rene $1000$ & $3.71$ & $0.02$ & $0.10$ & $0.31$ & $\mathbf{0.34}$ & $0.21$ & $0.03$ & \mbox{} \\


$2500$ &\rene $250$ & $3.30$ & $0.06$ & $0.14$ & $\mathbf{0.38}$ & $0.31$ & $0.12$ & \mbox{} & \mbox{} \\


$3000$ & \rene $250$ & $2.87$ & $0.11$ & $0.27$ & $\mathbf{0.35}$ & $0.19$ & $0.07$ & $0.01$ & \mbox{}   \\



$5000$ & \rene $155$ & $3.01$ & $0.14$ & $\mathbf{0.28}$ & $0.23$ & $0.15$ & $0.19$ & $0.01$ & \mbox{} \\

\end{tabular}
    \caption{
    Average number of domains $\overline{\mathbf{K}}$ and relative frequencies of structures with $\mathbf{K}$ domains in hexagonal cavities
    as in Tab.~\ref{tab_rel_freq_table}, but for systems with different particle numbers $N$ at area fraction $\eta$ = 0.75. The number of simulations from which the data are sampled is given in the second column.}
\label{tab_rel_freq_table_sizes}
\end{center}
\end{table}

\begin{table}[b]
\begin{center}
\begin{tabular}[t]{l|c|c|c|c|c|c|c|c|c|r}
 \mbox{} & \mbox{} & \mbox{} & \multicolumn{7}{c}{\textbf{number of domains $\mathbf{(K)}$}}&\\
 $\eta$ &  $p$ & $N$ &  $\overline{\mathbf{K}}$ & $\mathbf{1}$ & $\mathbf{2}$ & $\mathbf{3}$ & $\mathbf{4}$ & $\mathbf{5}$ & $\mathbf{6}$ & $\mathbf{7}$ \\
\hline 


$0.7$ & $10$ & $1147$ & $2.78$ & $0.09$ & $0.32$ & $\mathbf{0.38}$ & $0.16$ & $0.04$ & $0.01$    \\


\mbox{}  & $11.5$ & $1301$ & $2.49$ & $0.16$ & $\mathbf{0.37}$ & $0.32$ & $0.14$ & $0.02$ & \mbox{}  \\


\mbox{}  & $13$ & $1454$& $2.60$ & $0.12$ & $0.33$ & $\mathbf{0.39}$ & $0.14$ & $0.01$ & \mbox{}  \\


\mbox{}  & $15$ & $1659$ & $2.62$& $0.15$ & $0.28$ & $\mathbf{0.41}$ & $0.14$ & $0.03$ & \mbox{}  \\


\mbox{}  &  $18$ & $1965$ & $2.52$ & $0.14$ & $0.33$ & $\mathbf{0.41}$ & $0.11$ & $0.01$ & \mbox{}   \\

\hline


$0.725$ & $10$ & $1188$ & $3.18$ & $0.05$ & $0.20$ & $\mathbf{0.39}$ & $0.25$ & $0.10$ & $0.01$ &  \mbox{} \\


\mbox{}  &  $11.5$ & $1347$ & $3.09$ & $0.06$ & $0.20$ & $\mathbf{0.41}$ & $0.26$ & $0.06$ & $0.01$ &  \mbox{} \\


\mbox{}  &  $13$ & $1506$ & $3.30$ & $0.04$ & $0.16$ & $\mathbf{0.40}$ & $0.28$ & $0.11$ & $0.01$ &  \mbox{} \\


\mbox{}  &  $15$ & $1718$ & $3.23$ & $0.03$ & $0.20$ & $\mathbf{0.40}$ & $0.25$ & $0.11$ & $0.01$ &  \mbox{} \\


\mbox{}  & $18$ & $2036$ & $3.05$ & $0.06$ & $0.21$ & $\mathbf{0.41}$ & $0.25$ & $0.05$ & $0.01$ &  \mbox{} \\

\hline


$0.75$ & $10$ & $1229$ & $3.51$ & $0.03$ & $0.13$ & $\mathbf{0.35}$ & $0.30$ & $0.16$ & $0.03$ &  \mbox{} \\


\mbox{} & $11.5$ & $1394$ & $3.71$ & $0.01$ & $0.8$ & $0.33$ & $\mathbf{0.38}$ & $0.17$ & $0.02$ &  $0.01$ \\


\mbox{} & $13$ & $1558$ & $3.57$ & $0.03$ & $0.10$ & $\mathbf{0.35}$ & $0.34$ & $0.16$ & $0.02$ &  \mbox{} \\


\rowcolor{LightMagenta}
\mbox{} & $15$ & $1777$ & $3.74$ & $0.01$ & $0.12$ & $0.25$ & $\mathbf{0.38}$ & $0.22$ & $0.02$ &  \mbox{} \\


\mbox{} & $18$ & $2106$ & $3.56$ & $0.03$ & $0.13$ & $0.31$ & $\mathbf{0.35}$ & $0.17$ & $0.02$ & \mbox{} \\

\end{tabular}
    \caption{\rene
    Average number of domains $\overline{\mathbf{K}}$ and relative frequencies of structures with $\mathbf{K}$ domains in hexagonal cavities
    as in Tab.~\ref{tab_rel_freq_table}, but for different particle numbers $N$, area fractions $\eta$ and aspect ratios $p=L/D$, 
    fixing the ratio $7.5$ between side length of the confining hexagon and rod length.
    The number of simulations is equal to $300$ for each row. The parameters $\eta = 0.75$ and $p=15$ used in all other simulations are highlighted in magenta.}
\label{tab_rel_freq_table_eta_p}
\end{center}
\end{table}

\subsection{Polygons with different particle numbers} 
\label{sec_size}

The typical extent of the topological defects we observe in extreme confinement, depends on the elastic deformation energy of the smectic layers. On smaller length scales, positional distortions of the smectic layers are favored, whereas bent smectic layers are dominant on larger length scales \cite{SoftMPhys}. This results in the formation of the described networks of grain boundaries in smaller systems while in bigger systems large scale bends, connecting different walls of the confinement, suppress the formation of separated domains. 
Accordingly, we show in Figs.~\ref{fig_smaller_systems}, \ref{fig_imd_systems} and \ref{fig_bigger_systems} exemplary simulation results with different particle numbers $N$, {\rene leaving the area fraction $\eta=NA_\text{HDR}/\mathcal{A}=0.75$ constant for any confinement}. 
In general, we observe that the average number of domains in each confinement decreases with increasing system size, which is statistically quantified in Tab.~\ref{tab_rel_freq_table_sizes}.
 Moreover, the relative length of free-standing type-$B$ grain boundaries is reduced for larger systems,
which can be observed, e.g., by comparing the structures in the heptagons. 

The extremely confined structures with $N=500$ particles shown in \figref{fig_smaller_systems} display 
a tendency to develop pronounced positional distortions all across the central bridging domain. This is for instance visible through the slightly rotated dislocated layer in the hexagon and the strong elastic deformations in combination with the almost system-spanning central grain boundary in the heptagon. Moreover, the triangular cavity typically displays a single $\Qnem=1$ network featuring a positively charged node with $\Qtet=+1/4$, separating three equally-sized domains. Strictly speaking, this structure is different from the bridge state, irrespective of the identical set of building blocks.
For systems larger than those considered in the main text,
the enhanced flexibility leads to a higher frequency of smectic layers interrupting and thus disconnecting the individual grain-boundary networks. In return, this results in a higher frequency of structures with unified neighboring domains, as can be nicely observed for $N=2000$ in Fig.~\ref{fig_imd_systems}.
This behavior is exemplified in a more pronounced way for $N=5000$ in Fig.~\ref{fig_bigger_systems}, depicting a larger number {\rene of grain boundaries with partially annihilated tetratic end points}, while the length of all grain boundaries relative to the cavity size decreases. 

{\rene
\subsection{Comparison between simulation and experiment}
\label{sec_exp}

To establish a closer connection to the experimental data shown in the main manuscript,
we show in Tab.~\ref{tab_rel_freq_table_eta_p} a set of simulation data for systems with a constant ratio $a_\text{hex}/(L+D) = 7.5$ of the side length of the hexagonal confinement and the rod length.
This allows for the formation of seven to eight smectic layers at each side like in the experiment.
The remaining parameters are the area fraction $\eta$ and the rod aspect ratio $p$.

Keeping the packing fraction fixed, we find that the distribution of the different structures is largely independent of the aspect ratio, although a change in $p$ is accompanied by a relatively big change in the particle number $N$.
This means that the effect described in Sec.~\ref{sec_size} of increasing $N$ 
is mainly due to increasing the typical length scale of the confinement relative to the particle length.
Therefore, the difference of (effective) aspect ratios in our simulations and experiments has no significant effect.
In contrast, we observe a clear trend that the distribution of domain numbers
shifts to smaller values when the packing fraction $\eta$ is reduced.
In conclusion,  structures with more flexible smectic layers as well as less separated domains,
which are typical for our colloidal experiment, are best reproduced in our hard-rod simulations for 
 choosing a particle number $N$ and aspect ratio $p$ compatible with the desired number of layers in a given geometry
 and reducing the packing fraction $\eta$, see Fig.~\ref{fig_quantit_comp_pics} for an exemplary structure.

In principle, a closer match between simulation and experiment could likely be reached by taking into account
(i) the effects of the slight softness and polydispersity of the silica rods,
which also tend to reduce the rigidity of the smectic layers,
(ii) the three-dimensional nature of the experiment, which makes it difficult to estimate and compare an explicit packing fraction at the bottom,
and (iii) the particular equilibration protocol of the two-dimensional simulation,
contrasting the experimental sedimentation process
(also note that the compression in our experiments is faster than in Ref.~\cite{annulus}, since in this work we use a larger number of rods, which results in a higher osmotic pressure at the bottom of the cavity).}
%
%

\begin{figure}[t]
\begin{center}
\includegraphics[height=0.204\linewidth]{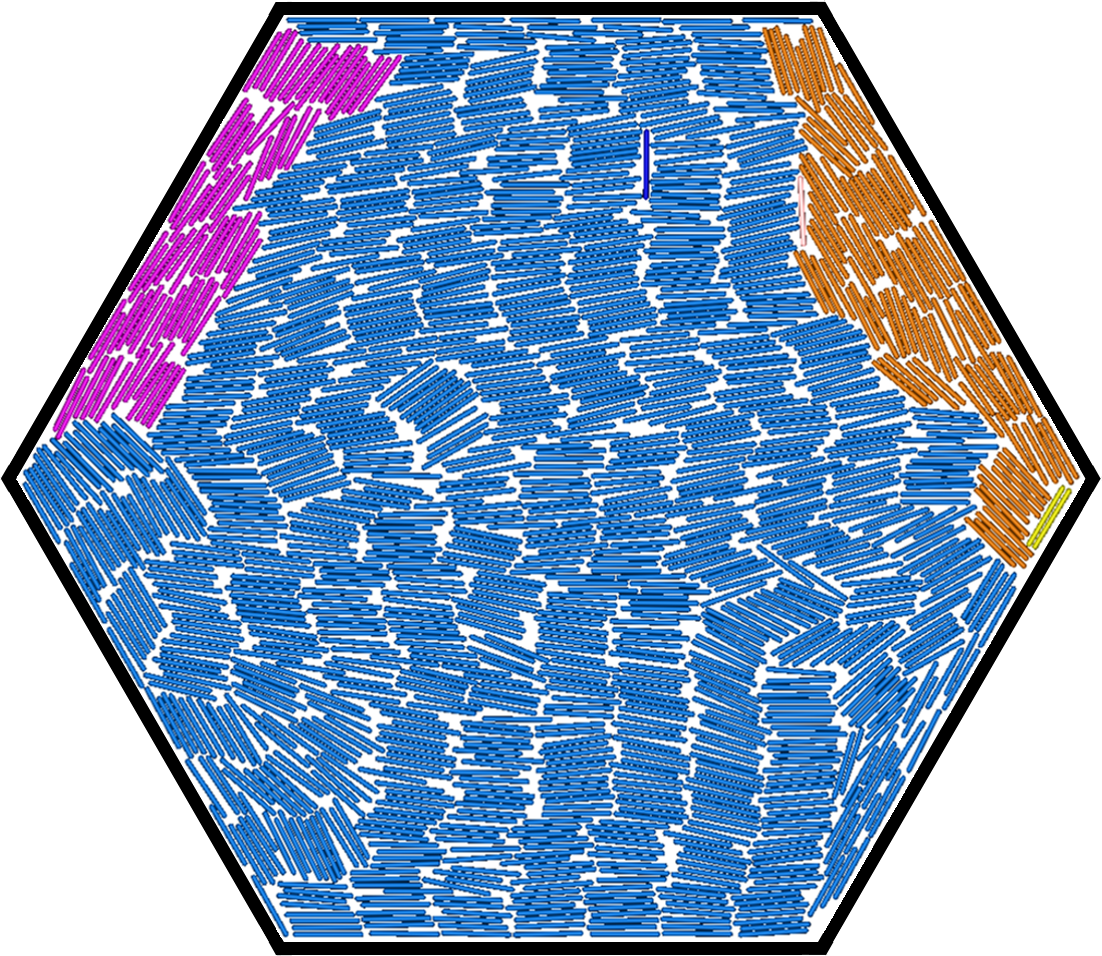}
~~~~~~
\includegraphics[height=0.204\linewidth]{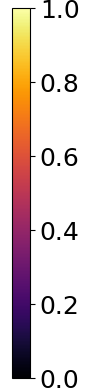}
\includegraphics[height=0.204\linewidth]{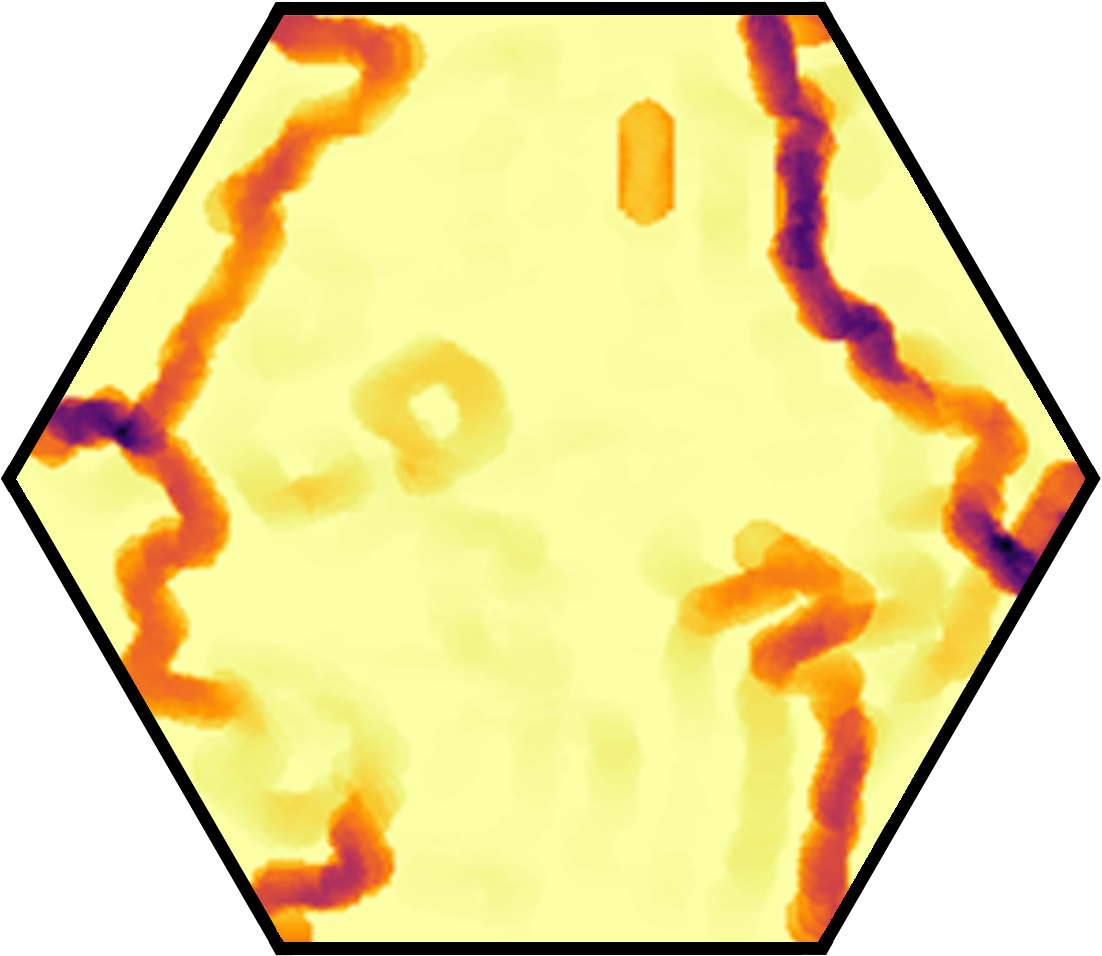}
~~~~~~
\includegraphics[height=0.204\linewidth]{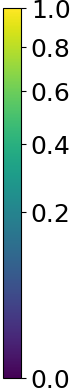}
\includegraphics[height=0.204\linewidth]{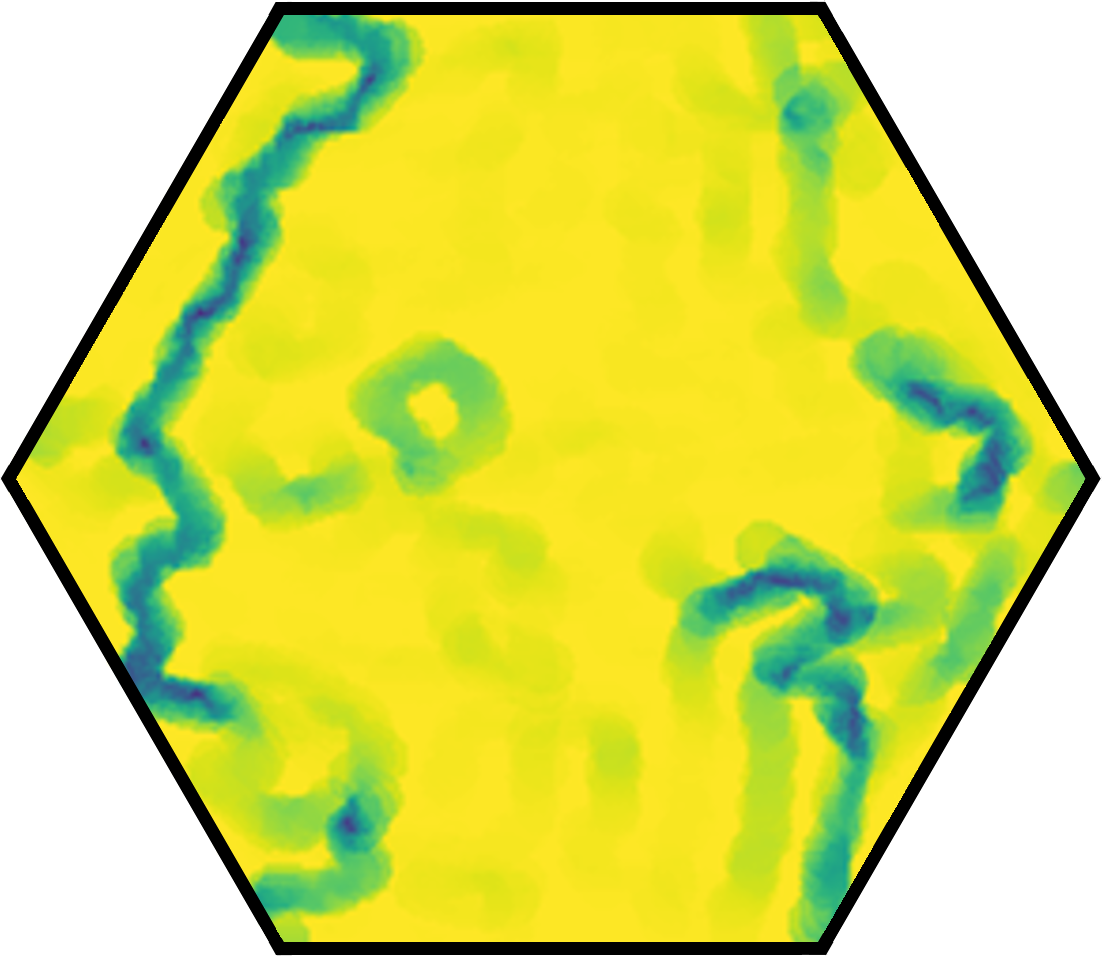}

\caption{\reneN Particle snapshot, nematic order parameter $S$ and tetratic order parameter $T$ for a representative simulation of hard rods in a hexagon with $N=1347$, $\eta=0.725$ and $p=11.5$. }
\label{fig_quantit_comp_pics}
\end{center}
\end{figure}

\end{widetext}

\end{document}